\documentclass[a4paper,fleqn,usenatbib]{mnras}

\usepackage{newtxtext,newtxmath}

\usepackage[T1]{fontenc}
\usepackage{ae,aecompl}

\usepackage{epsfig,graphicx}	
\usepackage{amsmath}	
\usepackage{amssymb}	
\usepackage{natbib}
\usepackage{bm}
\usepackage{comment}
\usepackage{dcolumn}
\usepackage{tabu}
\usepackage{datetime}
\settimeformat{hhmmsstime}
\usepackage{multicol}
\usepackage{pdflscape}
\usepackage{color}
\usepackage{needspace}

\newcommand\phn{\phantom{0}}%
\newcommand\arcdeg{\mbox{$^\circ$}}%

\title[Jet-ISM Interaction Zones]{
Mapping Jet-ISM Interactions in X-ray Binaries with ALMA: A GRS 1915+105 Case Study }

\author[A.J. Tetarenko et al.]{A.J. Tetarenko$^{1}$\thanks{E-mail: tetarenk@ualberta.ca},
P. Freeman$^{1,2}$,
E.W. Rosolowsky$^1$,
J.C.A Miller-Jones$^3$,
\newauthor
and G.R. Sivakoff$^1$
\\
$^{1}$Department of Physics, University of Alberta, CCIS 4-181, Edmonton, AB T6G 2E1, Canada\\
$^{2}$Department of Physics and Astronomy, University of Calgary, 2500 University Drive, Calgary, AB T2N 1N4, Canada\\
$^{3}$International Centre for Radio Astronomy Research -- Curtin University, GPO Box U1987, Perth, WA 6845, Australia\\
}
\date{Accepted XXX. Received YYY; in original form ZZZ}

\pubyear{2017}

\begin{document}
\label{firstpage}
\pagerange{\pageref{firstpage}--\pageref{lastpage}}
\maketitle

\begin{abstract}
We present Atacama Large Millimetre/Sub-Millimetre Array (ALMA) observations of IRAS 19132+1035, a candidate jet-ISM interaction zone near the black hole X-ray binary (BHXB) GRS 1915+105. With these ALMA observations (combining data from the 12 m array and the Atacama Compact Array), we map the molecular line emission 
across the IRAS 19132+1035 region. We detect emission from the $^{12}$CO [$J=2-1$], $^{13}$CO [$\nu=0$, $J=2-1$], C$^{18}$O [$J=2-1$], ${\rm H}_{2}{\rm CO}$ [$J=3_{0,3}-2_{0,2}$], ${\rm H}_{2}{\rm CO}$ [$J=3_{2,2}-2_{2,1}$], ${\rm H}_{2}{\rm CO}$ [$J=3_{2,1}-2_{2,0}$], SiO [$\nu=0$, $J=5-4$], CH$_3$OH [$J=4_{2,2}-3_{1,2}$], and CS [$\nu=0$, $J=5-4$] transitions. 
Given the morphological, spectral, and kinematic properties of this molecular emission, we present several lines of evidence that support the presence of a jet-ISM interaction at this site, including a jet-blown cavity in the molecular gas. {This compelling new evidence identifies this site as a jet-ISM interaction zone}, making GRS 1915$+$105 the third Galactic BHXB with at least one conclusive jet-ISM interaction zone. 
However, we find that this interaction occurs on much smaller scales than was postulated by previous work, where the BHXB jet does not appear to be dominantly powering the entire IRAS 19132+1035 region. 
Using estimates of the ISM conditions in the region, we utilize the detected cavity as a calorimeter to estimate the time-averaged power carried in the GRS 1915+105 jets of $(8.4^{+7.7}_{-8.1})\times10^{32}\,{\rm erg\,s}^{-1}$.
Overall, our analysis demonstrates that molecular lines are excellent diagnostic tools to identify and probe jet-ISM interaction zones near Galactic BHXBs.
\end{abstract}
\begin{keywords}
black hole physics --- ISM: individual objects: IRAS 19132$+$1035 --- ISM: jets and outflows --- stars: individual: GRS 1915+105 --- submillimeter: stars --- X-rays: binaries 
\end{keywords}



\section{Introduction}
Relativistic jets launched from accreting black holes carry a significant amount of energy and matter into their surrounding environment, and thus are important sources of galactic-scale feedback.
For instance, jets launched from super-massive black holes in Active Galactic Nuclei (AGN) are known to interact with the intergalactic medium (IGM) on cluster scales, carving out huge cavities in hot gas (e.g.\ \citealt{mac07}). These AGN jets are also thought to play a major role in galaxy formation and evolution (e.g.\ \citealt{mag98,cron06,mac05}). Similarly, the jets launched from Galactic black hole X-ray binaries (BHXBs), the stellar-mass analogues to AGN, also have an influence on their environment.
These objects release a significant portion of the liberated accretion power into their relativistic jets \citep{heigrimm,rus10}, injecting an estimated 1\% of the time-averaged luminosity of supernovae into the surrounding ISM \citep{fen05}.
This injected energy heats the ISM, generates interstellar turbulence, produces high-energy cosmic rays, seeds the ISM with magnetic fields, and possibly stimulates star formation \citep{heinz08,mir15}.

Models of the jet-ISM interaction in BHXBs (e.g.\ \citealt{kai04}) predict that as the jet impacts the ambient medium a strong radiative shock is likely to develop. Jet particles will inflate a radio lobe, which will expand to form a bubble of shock-compressed gas containing a population of relativistic electrons, producing non-thermal emission. However, we may not see all of these predicted features at every interaction site. Both the local environment and the BHXB properties may affect how these jet-ISM interactions manifest themselves (e.g.\ flux, morphology, chemistry).
For example, since most BHXBs are thought to propagate through a lower pressure and density environment (relative to jet power) when compared to AGN \citep{heinz02}, a local density enhancement in the surrounding medium is likely required for jet-blown lobes to form (e.g.\ in Cygnus X-1 the jet is believed to be moving through the tail of a dense H{\sc ii} region; \citealt{gallo05,rus07}). Further, 
jets launched from sources with lower peculiar velocities (relative to the local standard of rest) are more likely to inflate jet-blown lobes at the interaction site, as these sources 
have a more stable jet direction over time \citep{millerjones07}. On the other hand, sources with high peculiar velocities ($>100\,{\rm km\,s}^{-1}$) are more likely to produce trails of radio plasma, rather then a radio lobe structure, as the ram pressure of the ISM sweeps up the plasma released by the jet \citep{heinz08,wie09}.

Valuable information on unknown jet properties, most notably, the total jet power, radiative efficiency, jet speed, and the matter content are encoded within the properties of jet-ISM interaction regions (e.g.\ \citealt{mac07,bur59,cas75,heinz06}).
For instance, \cite{gallo05} used the jet-blown bubble as a calorimeter to estimate the total time-averaged power in the Cygnus X-1 jets to be $9\times 10^{35} \leq \, P_{\rm jet}\, \leq \, 10^{37} \, {\rm erg\,s}^{-1}$, by modelling the shell emission from the radio nebula as radiatively shocked gas.
Further work by \cite{rus07} and \cite{sell15} narrowed these estimates using a more tightly constrained shock velocity in the region.
These works clearly demonstrate that such calculations are highly sensitive to the properties of the shock and ISM (i.e.\ density, temperature, shock velocity). 
Therefore, placing improved observational constraints on these parameters in multiple jet-ISM interaction sites is crucial for such efforts.

As a jet-ISM interaction will create an environment with a unique chemistry, we expect to observe significant line emission from such a region.
In particular, molecular lines provide for excellent diagnostics of shock energetics and ISM excitation as different species can trace the density (CO), temperature (H$_2$CO; \citealt{gin15}), and presence of a shock in the gas (SiO, CS, and CH$_3$OH; \citealt{molecular} ). 
By mapping molecular line emission at potential jet impact sites near BHXBs, we can develop several lines of evidence to conclusively identify jet-ISM interaction regions, and accurately probe the ISM conditions at these sites.   

\begin{figure*}
\begin{center}
 \includegraphics[width=0.9\textwidth]{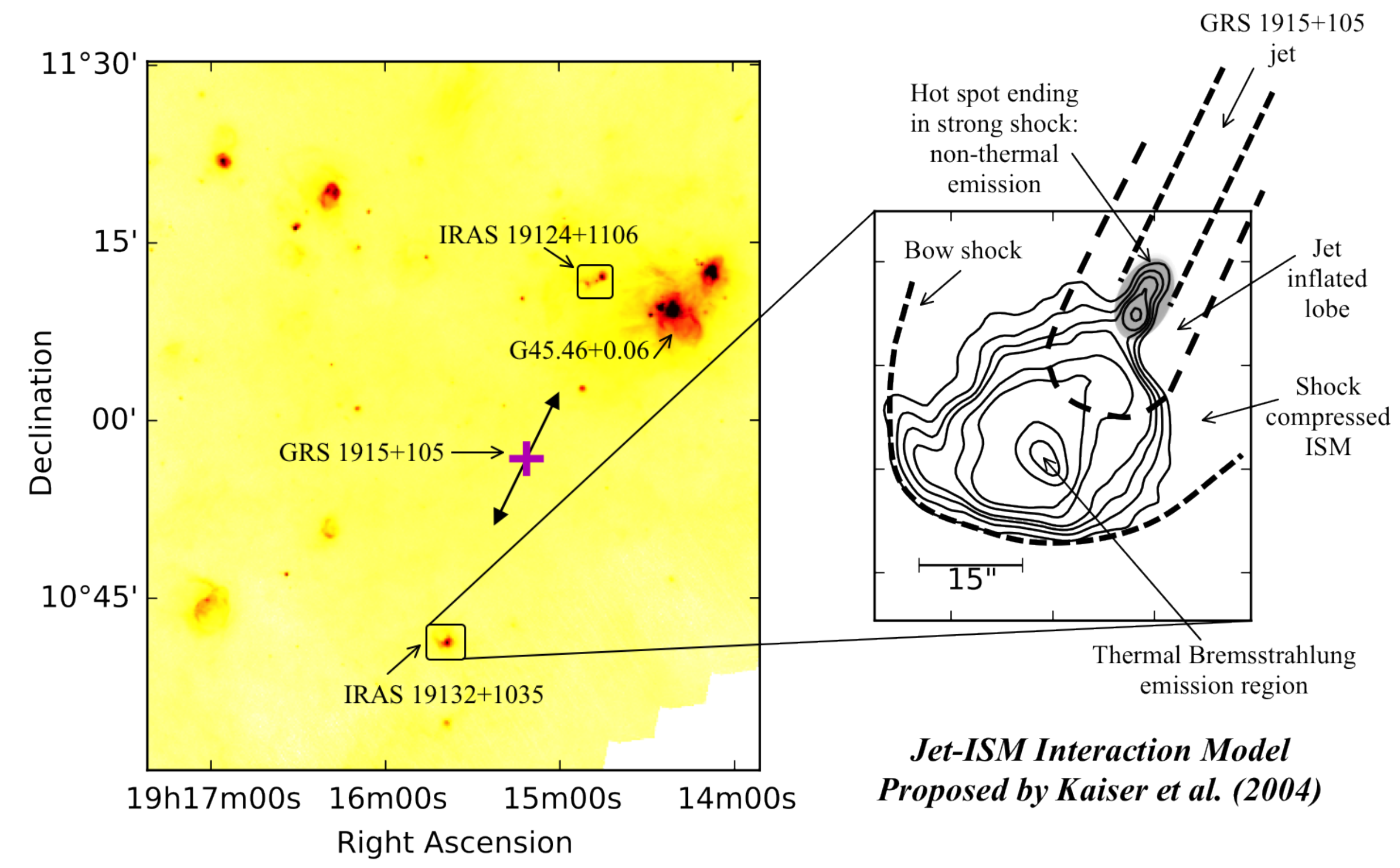}
 \caption{\small\label{fig:grs} (\textit{left}) A map of the region surrounding the BHXB GRS 1915$+$105, taken with the Herschel PACS continuum instrument at $70\mu m$. This map spans the extent of the previous VLA radio observations \citep{rodmir98}, 
 with which the two candidate interaction zones (IRAS 19132+1035 and IRAS 19124+1106) were first identified. Notable sources of emission are marked, including GRS 1915+105 (magenta plus sign), the position angle of the observed relativistic ejecta (solid-head arrows), the two candidate interaction zones (black squares), and a nearby compact H{\sc ii} region (G45.46+0.06). Our target source, IRAS 19132$+$1035, is southeast of GRS 1915$+$105. 
(\textit{right}) VLA (C-configuration) radio continuum map of IRAS 19132+1035 in the 4--8 GHz band, with a schematic sketch of the XRB jet-ISM interaction model (the schematic model is reproduced here to match Figure 2 in \citealt{kai04}). The contour levels are 0.2, 0.3, 0.4, 0.5, 0.75, 1, 2, and 3 ${\rm mJy\,bm}^{-1}$. The non-thermal radio jet feature is clearly labelled (shaded grey region), and dashed lines show model features that are not directly observed in the radio continuum image.
 }
 \end{center}
 \end{figure*}

To date, there are only two BHXBs (SS 433; \citealt{dubn98}, and Cygnus X-1; \citealt{gallo05}) where confirmed jet-ISM interaction sites have been detected (i.e.\ a jet-blown bubble/cavity and shock excited gas are observed).
However, a number of other potential jet-ISM interaction sites have been identified in the vicinity of BHXBs (1E1740--2942; \citealt{mir92}, GRS 1758-258; \citealt{marti02}, GRS 1915$+$105; \citealt{kai04,rodmir98,chat01}, H1743-322;, \citealt{co05}, XTE J1550-564; \citealt{cor02,karr3,migl17}, XTE J1748-288; \citealt{br7}, GRO J1655-40; \citealt{hjr95,hann00}, GX 339-4; \citealt{gall04}, 4U 1755-33; \citealt{kaa06}, XTE J1752-223; \citealt{yang10,millerjones11,yang11,ratti12}, XTE J1650-500; \citealt{corb04}, XTE J1908+094; \citealt{rush17}, and GRS 1009-45; \citealt{rus06b}) on the basis that their morphological or kinematic properties are consistent with models of jet-ISM interactions in BHXBs.
Out of all of these systems, GRS 1915+105 is an ideal candidate with which to study jet-ISM interactions through molecular tracers, as this system has been in a bright outburst period for over 25 years, ejects some of the most powerful relativistic jets in the known Galactic BHXB population \citep{fenbel04}, and its candidate interaction zones have existing molecular line detections \citep{chat01}. In this paper, we report on our Atacama Large Millimetre/Sub-Millimetre Array (ALMA) observations of the molecular line emission in IRAS 19132$+$1035, one of the candidate jet-ISM interaction zones near GRS 1915+105 (see Figure~\ref{fig:grs}).

 \subsection{GRS 1915+105}
GRS 1915+105 is a BHXB that was discovered in X-rays by the GRANAT satellite in 1992 \citep{ct92}.
The radio counterpart to the X-ray source was found shortly after with the Very Large Array (VLA; \citealt{mirbel3}). Follow up radio-frequency observations \citep{mirbel4} revealed resolved jet components, traveling away from the central source at apparent speeds that exceeded the speed of light, marking GRS 1915+105 as the first superluminal source discovered in the Galaxy.
Since its discovery, this source has remained in a bright outburst state. 

Recently, \cite{reid2014} obtained astrometric measurements of GRS 1915+105 with the  Very Long Baseline Array (VLBA), measuring a model-independent parallax distance of $8.6^{+2.0}_{-1.6}$ kpc, as well as proper motions
$\mu_{\rm ra} \cos\delta =-3.19\pm0.03$ ${\rm mas\, yr}^{-1}$ and $\mu_{\rm dec}=-6.24\pm0.05$ ${\rm mas\, yr}^{-1}$, which correspond to a peculiar motion\footnote{\cite{reid2014} use the convention that $U$ is towards Galactic centre at the location of GRS 1915+105, $V$ is in the direction of Galactic rotation, and $W$ is toward the north Galactic pole.} with respect to a circular Galactic orbit of ($U$,$V$,$W$)=($19\pm3$, $-10\pm24$, $6\pm2$) ${\rm km\, s}^{-1}$ (giving a total peculiar velocity of $22\pm24$\,${\rm km\, s}^{-1}$). Additionally, during one of their VLBA observations, \cite{reid2014} tracked the motion of a resolved jet component (travelling with a proper motion of $\mu=23.6\pm0.5$\,mas\,d$^{-1}$, resulting in a measured bulk speed of $0.81\pm0.04 c$), which when combined with their improved distance estimate, leads to an updated inclination angle measurement of $60\pm5^\circ$. 

 \subsection{Candidate interaction zones near GRS 1915+105}
\cite{rodmir98} identified two sources of bright radio emission, coincident with IRAS sources, in the vicinity of GRS 1915+105; IRAS 19124$+$1106 and IRAS 19132$+$1035 (see left panel of Figure~\ref{fig:grs}). Both sources are located $17$\arcmin\,away from GRS 1915+105 on the sky, which at a distance of 8.6 kpc, corresponds to $42.5$ pc. Based on their morphology and location, the authors proposed that these sources may be potential interaction zones between the GRS 1915+105 jet and the surrounding ISM. However, despite several observing campaigns (e.g.\ \citealt{rodmir98,chat01,millerjones07}), no definitive evidence has been presented that confirms that 
these IRAS sources originated as a result of (or are dominantly powered by) jet-ISM interactions.
There is a plethora of circumstantial evidence both for and against IRAS 19132+1035 being a jet-ISM interaction: 
this source shares the same position angle as jet ejections from GRS 1915$+$105 observed at radio wavelengths ($\sim$130--151$^\circ$, see Table 2 in \citealt{reid2014});
there is a non-thermal linear radio emission feature that is spatially coincident with the inner edge of IRAS 19132$+$1035 and aligned with the jet axis in GRS 1915$+$105 (see right panel of Figure~\ref{fig:grs});
the highest densities in IRAS 19132$+$1035 are located on the side nearest to the central BHXB \citep{chat01};
and a bow-shock like structure may be located on the side of IRAS 19132$+$1035 farthest from the central BHXB \citep{rodmir98}. Furthermore, the recent parallax distance determination to GRS 1915$+$105 ($8.6^{+2.0}_{-1.6}$ kpc; \citealt{reid2014}) brings this BHXB closer to the inferred distance of IRAS 19132$+$1035 ($6.0\pm1.4$ kpc\footnote{This is a kinematic distance,
while the GRS 1915+105 distance is a model-independent geometric parallax distance. As kinematic distances are known to be less accurate than parallax distances \citep{reid2014b}, for all remaining calculations involving distance we will use the parallax distance, but see \S\ref{sec:dist} for a discussion on the effect of distance estimates on our results.}; \citealt{rodmir98}). 
On the other hand, no high-velocity shock feature was seen in previous spectral line data \citep{chat01}, no high-energy X-ray emission was detected at the suspected impact site \citep{millerjones07}, and the luminosity and morphology of the region are consistent with a high mass star forming region, dominantly powered by one or more hot stars \citep{rodmir98}.

 \renewcommand\tabcolsep{7pt}
 \begin{table*}
\caption{ALMA Correlator Setup}\quad
\centering
\begin{tabular}{ llcccc }
 \hline\hline
 {\bf Base-band}& {\bf Target Line}&{\bf Central Sky}&{\bf Bandwidth}&{\bf Resolution}&{\bf Number}\\
  {\bf }& {\bf }&{\bf Frequency}&{\bf  (${\frac{\rm \bf km}{\rm \bf s}}$ / MHz)}&{\bf  (${\frac{\rm \bf km}{\rm \bf s}}$ / kHz)}&{\bf {of}}\\[0.15cm]
    {\bf }& {\bf }&{\bf (GHz)}&{\bf }&{\bf }&{\bf Channels}\\[0.15cm]
  \hline
1& $^{12}$CO ($J=2-1$)&230.48671&152 / 117.188&0.315 / 242.310&480\\[0.1cm]
1& N$_2$D+ ($J=3-2$)&231.27037&152 / 117.188&0.314 / 242.310&480\\[0.1cm]
2&continuum / CS ($\nu=0$, $J=5-4$)&231.84934&2424 / 1875 &2.507 / 1938&960\\[0.1cm]
3&SiO ($\nu=0$, $J=5-4$)&217.05668&324 / 234.375&0.674 / 488.281&960\\[0.1cm]
3&${\rm H}_{2}{\rm CO}$ ($J=3_{0,3}-2_{0,2}$)&218.17365&322 / 234.375&0.671 / 488.281&960\\[0.1cm]
3&${\rm H}_{2}{\rm CO}$ ($J=3_{2,2}-2_{2,1}$) / CH$_3$OH ($J=4_{2,2}-3_{1,2}$)&218.42703&322 / 234.375&0.670 / 488.281&960\\[0.1cm]
3&${\rm H}_{2}{\rm CO}$ ($J=3_{2,1}-2_{2,0}$)&218.71140&321 / 234.375&0.669 / 488.281&960\\[0.1cm]
4&C$^{18}$O ($J=2-1$)&219.51151&160 / 117.188&0.331 / 242.310&480\\[0.1cm]
4&$^{13}$CO ($\nu=0$, $J=2-1$)&220.34965&159 / 117.188&0.330 / 242.310&480\\[0.15cm] \hline
\end{tabular}\\
\begin{flushleft}
\end{flushleft}
\label{table:almasetup}
\end{table*}
\renewcommand\tabcolsep{6pt}

To resolve this long-standing question we obtained ALMA observations to map the molecular line emission in IRAS 19132+1035, constrain ISM properties in this region, and in turn determine if there is sufficient evidence to call IRAS 19132+1035 the third conclusive site of a BHXB jet-ISM interaction in our Galaxy.
In \S\ref{sec:oda} we describe the data collection and reduction process. In \S\ref{sec:imaging} we describe our custom imaging procedure. In \S\ref{sec:res} we present maps of the radio/sub-mm continuum and molecular line emission in the IRAS 19132$+$1035 region (density tracer, CO; temperature tracer, H$_2$CO; and shock tracers, SiO, CS, CH$_3$OH, N$_2$D$+$), outline the morphological, spectral, and kinematic properties of this emission, and present constraints on the temperature, density, and column density across the region.
In \S\ref{sec:discuss}, we discuss the ISM conditions in the IRAS 19132+1035 region, and what these conditions reveal about the presence of a jet-ISM interaction at this site.
We also present a comparison between IRAS 19132$+$1035, and other jet-ISM interaction zones. A summary of our work is presented in \S\ref{sec:sum}.

\section{Observations and Data Analysis}
\label{sec:oda}
\subsection{ALMA sub-mm observations}
We observed  IRAS 19132$+$1035 (Project Code: 2015.1.00976.S, PI: A.~Tetarenko) using the ALMA 12 m array (2016 Jan 14), as well as the Atacama Compact Array (ACA) 7 m (2016 May 13, 22, and 23) and total power arrays (executions between 2016 Mar 21--2016 April 16), with the Band 6 receiver (211--275 GHz)\footnote{Although GRS 1915$+$105 is a variable source, the evidence for jet-ISM interactions that we are exploring is not strongly variable on the timescales separating the different ALMA array observations.}.
During our observations, the 12 m array was in the most compact C36-1 configuration (with 43 antennas), and spent 34.2 min on source. We observed a 75\arcsec $\times$ 60\arcsec rectangular field centred on the coordinates (J2000) RA 19:15:39.1300, Dec 10:41:17.100 (peak of the radio continuum measured from previous VLA observations; Table 1 in \citealt{chat01}), which consisted of 33 pointings with the 12 m array and 14 pointings with the 7 m array (see Appendix~\ref{sec:appen} for details on our choice of mosaic field). The correlator was set up to yield $4\times2$ GHz wide base-bands, within which we defined 9 individual spectral windows centred  on our target molecular lines (see Table~\ref{table:almasetup} for the central frequencies, bandwidth, and resolution of these spectral windows).
All of the data were reduced and imaged (see \S\ref{sec:imaging} for imaging details) within the Common Astronomy Software Application package (CASA, version 4.7.2; \citealt{mc07}).
Flagging and calibration of the 12 m and total power data were performed with the ALMA pipelines, while flagging and calibration of the 7 m data were performed manually using standard procedures.
For the 12 m array, we used J1751+0939 as a bandpass calibrator, Pallas as a flux calibrator, and J1922+1530 as a phase calibrator.
For the 7 m array, we used J1924-2914 as a bandpass calibrator, J1751+0939 as a flux calibrator, and J1922+1530 as a phase calibrator.

\subsection{VLA radio continuum observations}
We downloaded and reduced public archival VLA observations of IRAS 19132$+$1035 (Project Code: 14B-482, PI: F.~Mirabel). These observations were taken on 2014 Dec 06, and consisted of scans on source from 22:02:34.0--22:57:52.0 UTC (MJD 56997.9184--56997.9569), in the C ($4$--$8\,{\rm GHz}$) band. The array was in its C configuration during the observations. All observations were made with a 3-bit sampler, comprised of 2 base-bands, each with 16 spectral windows of $64\times2$ MHz channels, giving a total bandwidth of 2.048 GHz per base-band. Flagging, calibration, and imaging of the data were carried out within CASA using standard procedures. J0137+3309 was used as a flux calibrator, and J1922+1530 was used as a phase calibrator. No self-calibration was performed. We imaged the source with two Taylor terms to account for the wide bandwidth, Briggs weighting with a robust parameter of 0.5 to balance sensitivity and angular resolution, and the multi-scale clean algorithm (scales of $[0,5,20,50]\times$ the pixel size of 0.7\arcsec) to effectively deconvolve extended emission. Flux densities from these observations are reported in Table~\ref{table:mwemiss}, and are discussed in \S\ref{sec:cont}. 

\subsection{Other multi-wavelength observations}
We compiled observations of the IRAS 19132$+$1035 region at other wavelengths (spanning $1.11$mm -- $2.73\mu$m), taken with various instruments/surveys, including CSO BOLOCAM (1.11 mm; \citealt{gin13}), Herschel SPIRE (500, 350, 250 $\mu$m; \citealt{mol10,grif10}), Herschel PACS (70 $\mu$m; \citealt{pog10}), NASA WISE (22 $\mu$m; \citealt{wri10}), Spitzer GLIMPSE (8.0 $\mu$m; \citealt{ben03,church09}), and UKIRT UKIDSS (2.37$\mu$m; \citealt{law07,casa07}). Flux densities from these observations are reported in Table~\ref{table:mwemiss}, and are discussed in \S\ref{sec:cont}.

\renewcommand\tabcolsep{3pt}
 \begin{table}
\caption{Continuum Emission Properties of IRAS 19132+1035}\quad
\centering
\begin{tabular}{ lcccc }
 \hline\hline
 {\bf Band}&{\bf $\bm{\nu}$}&{\bf $\bm{\lambda}$}&{\bf $\mathbf{F_{\rm \bf peak}}$}&{\bf $\mathbf{F_{\rm \bf integ}}$}\\
{\bf}& {\bf (GHz)}&{\bf ($\mathbf{\mu m}$)}&{\bf ($\mathbf{{\rm \bf mJy\, bm}^{-1}}$)$^c$}&{\bf (mJy)}\\[0.15cm]
  \hline
Radio&6.0&5e4&$2.19\pm0.11$&$ 64.8\pm3.3$\\[0.1cm]
Radio NT$^a$&6.0&5e4&$0.66\pm0.05$&$3.8\pm0.4$\\[0.1cm]
Sub-mm$^b$&232.0&1290&$1.23\pm0.15$&$ 117\pm15$\\[0.1cm]
Sub-mm&270.1&1110&$554\pm31$&$ (1.7 \pm 0.1)\times10^3 $\\[0.1cm]
mid-IR&599.6&500&$(10.4\pm0.1)\times10^{3}$&$ (42.3 \pm 0.7)\times10^3 $\\[0.1cm]
mid-IR&856.5&350&$(17.6\pm0.4)\times10^{3}$&$ (87.8 \pm 2.6)\times10^3 $\\[0.1cm]
mid-IR&1199.2&250&$(26.1\pm1.1)\times10^{3}$&$ (168.5 \pm 8.4)\times10^3 $\\[0.1cm]
mid-IR&4282.0&70&$(21.8\pm1.9)\times10^{3}$&$ (186 \pm 18)\times10^3 $\\[0.1cm]
mid-IR&1.4e4&22&$0.48\pm0.02$&$ 6.4 \pm 0.2 $\\[0.1cm]
NIR/Opt&3.7e4&8&$13.8\pm0.8$&$ 69.1\pm 4.9$\\[0.1cm]
Opt/UV&1.1e5&2.73&$64.9\pm5.3$&$ (1.5 \pm 0.1)\times10^3 $\\[0.1cm]

\hline
\end{tabular}\\
\begin{flushleft}
{$^a$ Fluxes reported in this row are measured over the linear non-thermal emission region only (see right panel of Figure~\ref{fig:grs}).}\\
{$^b$ Fluxes reported in this row are measured from our new ALMA observations.}\\
{$^c$ The approximate beam sizes for these images are 3.7, 1.4, 33, 37, 25, 18, 6.5, 10.8, 2.5, and 0.4\arcsec at 5e4, 1290, 1110, 500, 350, 250, 70, 22, 8, and 2.73 $\mu m$.}
\end{flushleft} 
\label{table:mwemiss}
\end{table}
\renewcommand\tabcolsep{6pt}

\section{ALMA imaging process}
\label{sec:imaging}
\subsection{Spectral line imaging}
We use a custom procedure, which combines the 12 m, 7 m, and total power array data, to image all of our spectral lines. First, we split out the respective spectral windows of the line being imaged in the 12 m and 7 m data, using the CASA \texttt{split} task. Since the 7 m data were taken in 3 different executions, with the spectral windows slightly offset in frequency between executions, we use the \texttt{mstransform} task to combine the 7 m spectral windows, for each line, from each execution. We then create a template 12 m + 7 m image by running the \texttt{clean} task with both of the 12 m and 7 m split measurement sets as input, and a single CLEAN iteration. The accompanying single dish image for the spectral window (i.e.\ the total power array image output from the ALMA pipeline) is re-gridded to this template image using the \texttt{imregrid} task. We ensure that this re-gridded single dish image has the same primary beam response (i.e.\ lower in the outskirts of the image) as the 12 m + 7 m  template by multiplying the re-gridded image by the primary beam template image (i.e.\ *.flux image in CASA).
Using the \texttt{imtrans} task , we rearrange the stokes axis of the re-gridded image to be compatible with the CASA \texttt{clean} task. We then run the \texttt{clean} task with both of the 12 m and 7 m split measurement sets as input, and the re-gridded single dish image as a model image. Following this, we take the positive-only interferometer components from the CLEAN model, smooth them to the synthesized beam using the \texttt{imsmooth} task, and combine the smoothed interferometer component image with the re-gridded single dish image, using the \texttt{feather} task. Lastly, we run the \texttt{clean} task again with both of the 12 m and 7 m split measurement sets as input, and the feathered single dish image as a model image.
All imaging is done with natural weighting to maximize sensitivity, a pixel size of 0.2 arcsec, and the multi-scale algorithm (scales of $[0,1,5,10,15]\times$ the pixel size). 
See Appendix A (Figure~\ref{fig:pbnoise}) for a primary beam noise map of the region. {We also tested the combination of single-dish and interferometer data directly through weighted averaging in the Fourier plane using the \texttt{feather} task in {\sc casa} only, and did not find any clear improvements in image quality.}

\begin{figure}
\begin{center}
 \includegraphics[width=1\columnwidth]{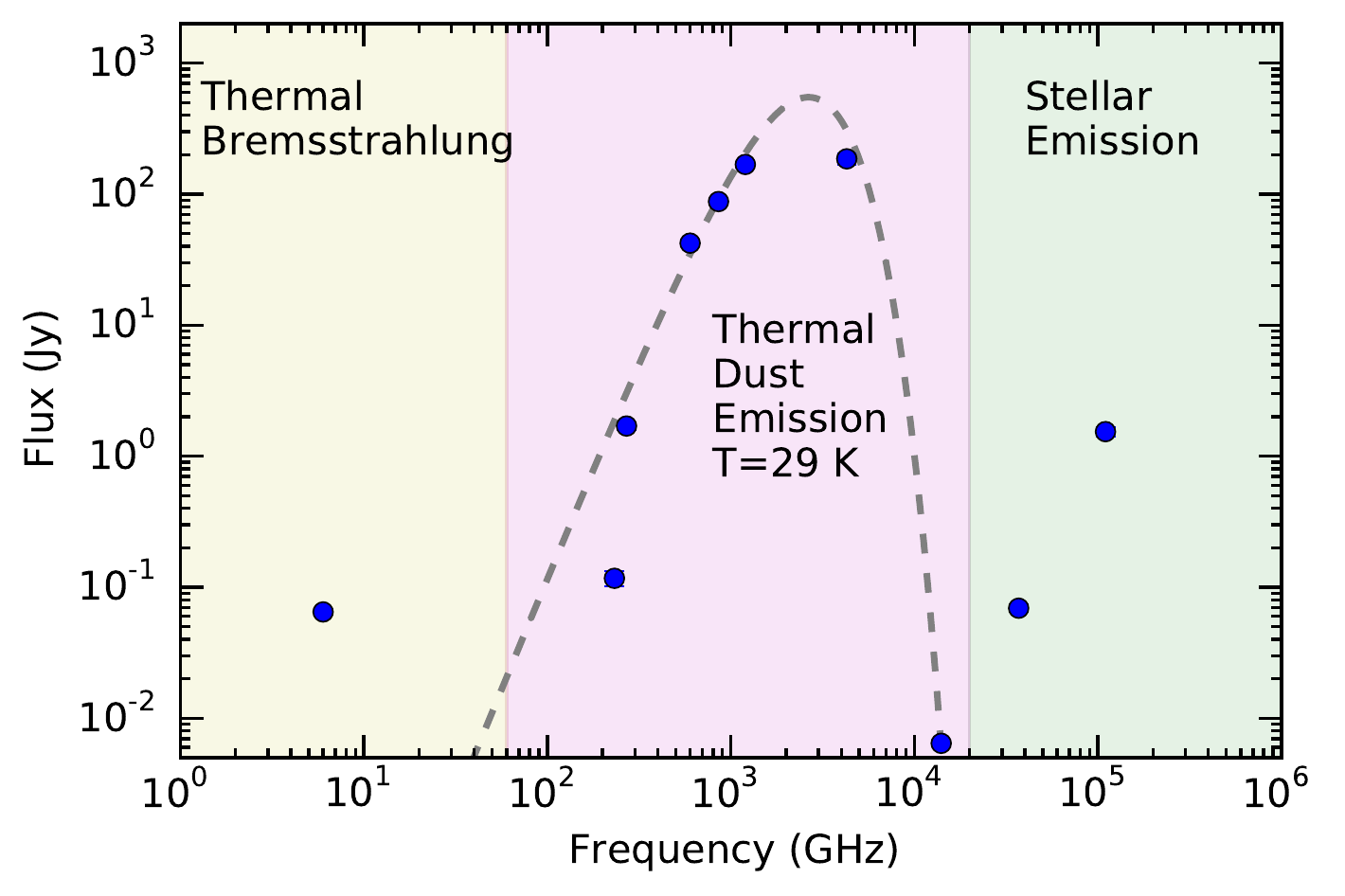}
 \caption{\small \label{fig:contSED} Broad-band spectrum of the continuum emission from IRAS 19132+1035. The blue points indicate the integrated flux densities from Table~\ref{table:mwemiss}, and the grey dotted line indicates a modified blackbody fit to the 22--1110\,$\mu$m data points (see text in \S\ref{sec:cont} for details). Different emission processes (highlighted colour bands labeled on plot) dominate the IRAS 19132+1035 continuum emission in different frequency ranges.}
 \end{center}
 \end{figure}

\renewcommand\tabcolsep{6pt}
\begin{figure*}
\begin{center}
 \includegraphics[width=1.9\columnwidth]{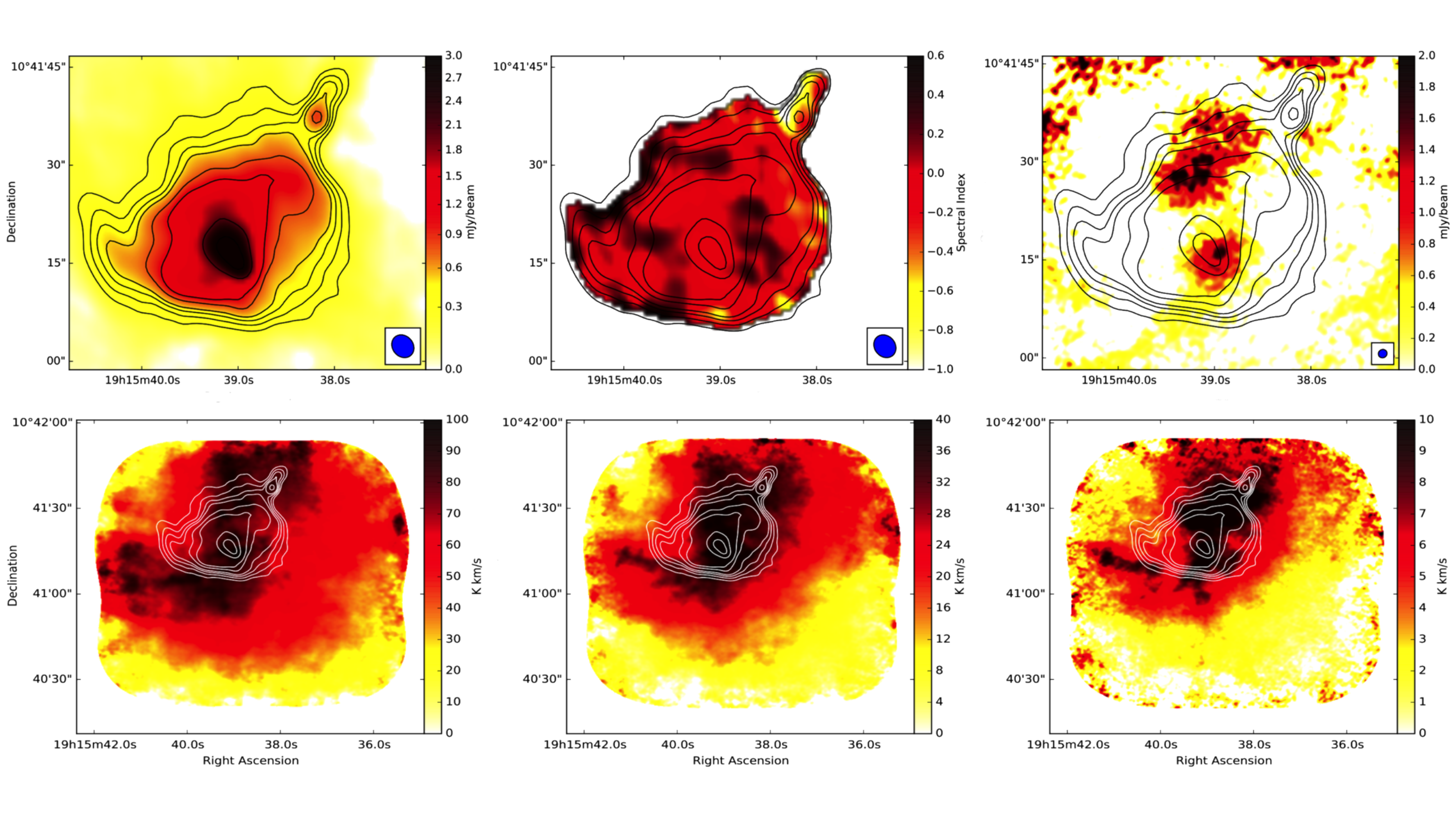}
 \caption{\small \label{fig:grsconts} Radio through sub-mm maps of the IRAS 19132$+$1035 region. The {\it top row, \it left to \it right}, displays the VLA continuum radio frequency image in the 4--8~GHz band (contour levels of 0.2, 0.3, 0.4, 0.5, 0.75, 1, 2, and 3 ${\rm mJy\,bm}^{-1}$), the radio spectral index map (using the convention $f_\nu\propto\nu^\alpha$, where $\alpha$ represents the spectral index), and the ALMA sub-mm continuum image in the 231.9--232.8 GHz band. The blue ellipses in the bottom right corners represent the VLA and ALMA beams. The {\it bottom row, \it left to right}, displays the $^{12}$CO, $^{13}$CO, and C$^{18}$O, integrated intensity maps; contours are the VLA radio frequency contours. The colour bars indicate the flux density in units of ${\rm mJy\,bm}^{-1}$ ({\it top left panel} \& {\it top right panels}), spectral index ({\it top middle panel}), or integrated intensity in units of ${\rm K\,km\,s}^{-1}$ ({\it bottom panels}). In the IRAS 19132+1035 region, the location of bright molecular emission coincides with that of the continuum emission.}
 \end{center}
 \end{figure*}

\begin{figure*}
\centering
\includegraphics[width=1.5\columnwidth]{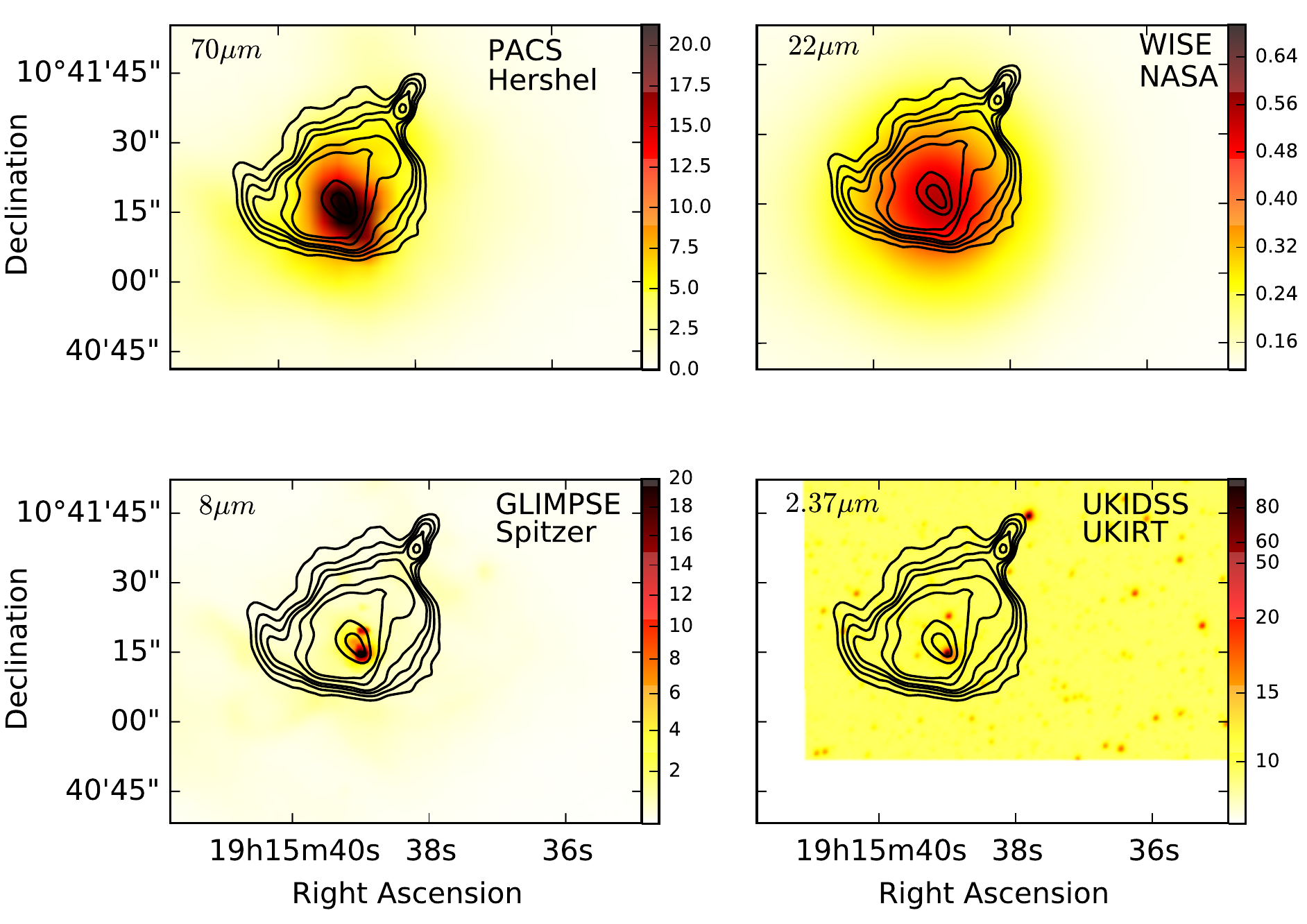}
\caption{\small \label{fig:grsconts_mw} Multi-wavelength continuum observations of the IRAS 19132$+$1035 region. Each panel is labeled with the wavelength (top left), and the instrument/telescope/survey (top right) for that specific image. The contours in all panels are the VLA radio frequency contours, as seen in Figure~\ref{fig:grsconts}; 0.2, 0.3, 0.4, 0.5, 0.75, 1, 2, and 3 ${\rm mJy\,bm}^{-1}$. The colour bars indicate the flux density in units of ${\rm mJy\,bm}^{-1}$, except for the {\it top left panel} which has units of ${\rm Jy\,bm}^{-1}$ . These multi-wavelength images display the extent of the contributions from other continuum emission sources in the region, namely dust ({\it top row}) and stellar emission ({\it bottom row}). }
\end{figure*}

\subsection{Continuum imaging}
To image the ALMA continuum emission, we split out the continuum spectral window in the 12 m and 7 m data, using the CASA \texttt{split} task, and then used the \texttt{mstransform} task to combine the 7 m continuum spectral windows from each execution. We then flag the channels with clear line emission in this spectral window (determined by examining the data in \texttt{plotms}), and use the \texttt{clean} task to perform multi-frequency synthesis imaging on the combined 12 m + 7 m data.
In this imaging process, we used natural weighting to maximize sensitivity, and two Taylor terms to account for the wider bandwidth. Note that the single dish data from the total power array does not have a continuum component.

\section{Results}
\label{sec:res}
\subsection{Continuum Emission from IRAS 19132+1035}
\label{sec:cont}

The top panels of Figure~\ref{fig:grsconts} display radio continuum, radio spectral index, and sub-mm continuum emission maps of the IRAS 19132+1035 region;  Figure~\ref{fig:grsconts_mw} displays IR through UV continuum emission from the IRAS 19132+1035 region.

The radio continuum emission is mainly extended over a 43\arcsec $\times$ 41\arcsec region, and 
displays a unique morphology; a linear feature (dimensions of $8$\arcsec $\times$ $11$\arcsec) extending in the direction of the central BHXB, and a sharp edge on the opposite side of the region.
The emission from the linear feature shows a steep spectral index ($\alpha\sim-0.7$, where $f_\nu\propto\nu^\alpha$) that is distinct from the flat ($\alpha\sim0$) spectral index observed in the rest of the region. 
This suggests that the emission in these two zones originates from different processes. In particular, the steep spectral index is consistent with non-thermal synchrotron emission from a relativistic plasma, commonly observed from BHXB jets \citep{fend06}, while the flat-spectrum emission is consistent with thermal bremsstrahlung emission from ionized hydrogen gas. 
Previous work \citep{rodmir98,kai04} estimated the temperature of the ionized hydrogen gas, using the width of a hydrogen recombination line (H92$\alpha$) detected in this region, to be $1.2\times10^4$ K.
The sharp edge seen in the radio continuum emission, on the side of the IRAS 19132+1035 region furthest from the central BHXB, has previously been associated with a potential bow-shock feature created by a jet-ISM interaction, or the ionization front from an H{\sc ii} region \citep{chat01}.

At sub-mm frequencies, the continuum emission in IRAS 19132+1035 
is confined to two regions, which are much more compact when compared with the extent of the radio continuum feature. One of these sub-mm continuum regions is consistent with the location of the peak flux density in the radio continuum (dimensions of $\sim 11\,\arcsec \times13\,\arcsec$), and the other is located to the north of the radio continuum peak (dimensions of $\sim 19\,\arcsec\times18\,\arcsec$). We do not detect sub-mm emission from the linear radio jet feature. This non-detection is expected, given that an extrapolation of the steep synchrotron spectrum to sub-mm frequencies predicts a flux density ($\sim54\mu$Jy) below our ALMA detection limits ($\sim130\mu$Jy).

\renewcommand\tabcolsep{1pt}
 \begin{table}
\caption{Best-fit parameters for thermal dust emission in IRAS 19132+1035}\quad
\centering
\begin{tabular}{ lp{12pt}rl }
 \hline\hline
 {\bf Parameter}&&\multicolumn{2}{c}{\bf Best-fit value}\\[0.15cm]
  \hline
$T_{\rm dust}$&&$29.01^{+0.14}_{-0.13}$& K\\[0.15cm]
$\Sigma_{\rm mol}$&&$450.0^{+11.1}_{-11.0}$& $M_\odot\,{\rm pc}^{-2}$\\[0.15cm]
$\beta$&&$1.44^{+0.03}_{-0.03}$&\\[0.15cm]
$M_{\mathrm{mol}}^a$&&$1200^{+690}_{-450}$& $M_\odot$\\[0.15cm]
\hline
\end{tabular}\\
\begin{flushleft}
{$^a$ The mass of the molecular medium derived from the dust emission is not a fitted parameter, but found using the expression $M_{\rm mol}=\Sigma_{\rm mol}\Omega D^2$. To properly account for uncertainties in distance (D) and surface density ($\Sigma_{\rm mol}$) we performed Monte Carlo simulations, sampling from the posterior distribution of $\Sigma_{\rm mol}$ output from the fitting procedure, and the known geometric parallax distance distribution \citep{reid2014}}.
\end{flushleft}
\label{table:greybody}
\end{table}
\renewcommand\tabcolsep{6pt}

\begin{figure*}
\begin{center}
\includegraphics[width=1.7\columnwidth]{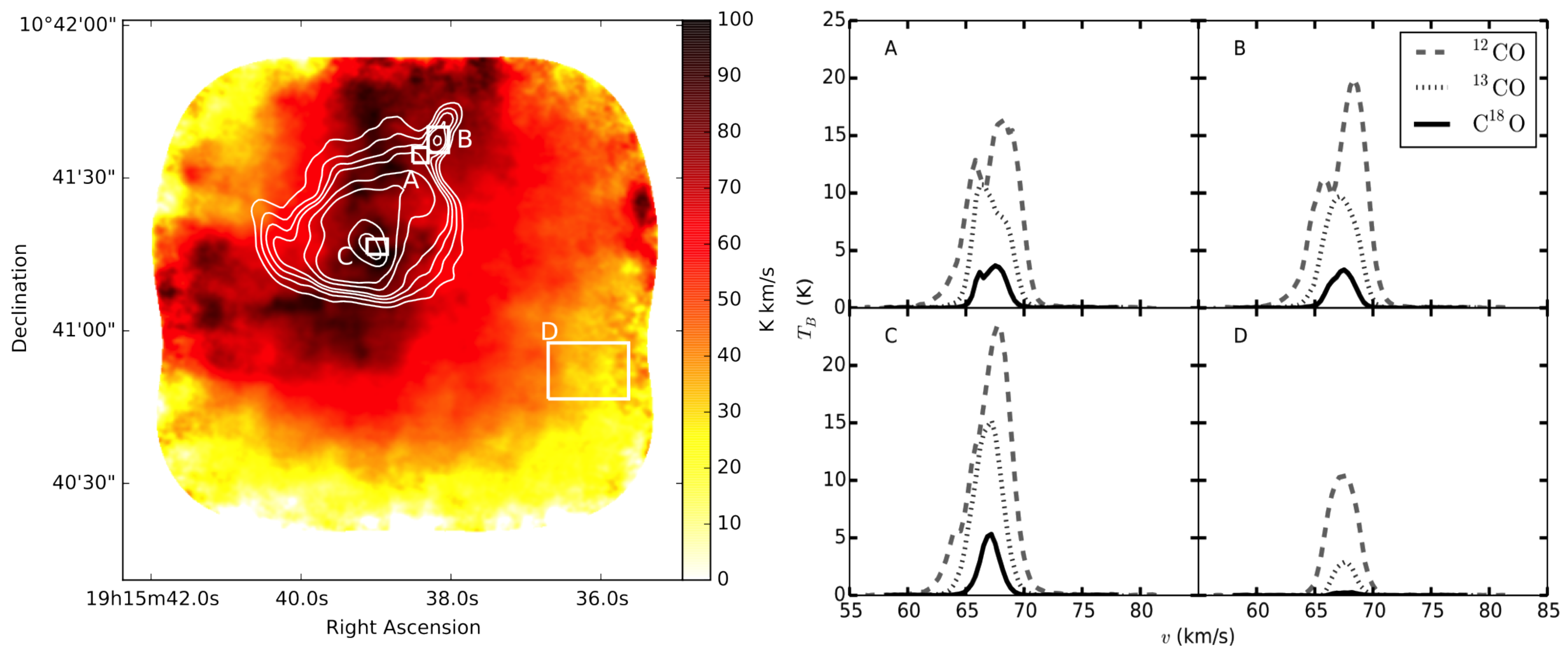}
\caption{\small \label{fig:CO_mom0_spec} CO emission in different zones of the IRAS 19132$+$1035 region. The {\it left panel} displays the $^{12}$CO integrated intensity map with four marked regions. The contours are the VLA radio frequency contours; levels of 0.2, 0.3, 0.4, 0.5, 0.75, 1, 2, and 3 ${\rm mJy\,bm}^{-1}$ (see Figure~\ref{fig:grsconts}). The {\it right panel} displays the spectra of the different isotopologues of CO taken in the labelled regions; $^{12}$CO (dashed grey line), $^{13}$CO (dotted light grey line), and C$^{18}$O (solid black line). The CO line profiles and peak intensities clearly vary between different zones in the IRAS 19132+1035 region. }
\end{center}
\end{figure*}

The bulk of the emission at mid-IR frequencies (i.e.\ $22$--$500\mu m$) in IRAS 19132+1035 is extended, and centred  on the thermal radio continuum feature. Given the integrated fluxes reported in Table~\ref{table:mwemiss}, both the sub-mm and IR continuum emission likely originate from thermal dust emission. This thermal dust emission is best modelled as a modified black-body, represented by,

\begin{equation}
\label{eq:modbb}
I_\nu=\frac{2h\nu^3}{c^2}\frac{\kappa_\nu\Sigma_{\rm mol}}{\left({\rm exp}\left[\frac{h\nu}{kT_{\rm dust}}\right]-1\right)}
\end{equation}
where we use the opacity law \citep{beck90},
\begin{equation}\nonumber
\kappa_{\nu} = 0.1~\mathrm{cm^2~g^{-1}}, (\nu/\mathrm{1~THz})^\beta,
\end{equation}
$\beta$ represents the emissivity index, $T_{\rm dust}$ represents the dust temperature, and $\Sigma_{\rm mol}$ represents the molecular gas surface density implied by the dust emission.  Note that this value of the dust opacity includes a gas-to-dust ratio of 100 by mass.

To estimate the physical properties of the dust emission region, we fit (using a Markov Chain Monte Carlo (MCMC) algorithm) the broad-band sub-mm through IR (22--1110\,$\mu$m; see Table~\ref{table:mwemiss}) spectrum, with the modified black-body function of Equation~\ref{eq:modbb}.
The best-fit parameters are displayed in Table~\ref{table:greybody}, and the broad-band spectrum (with the best-fit dust model over-plotted), is shown in Figure~\ref{fig:contSED}. We do not include our ALMA sub-mm continuum flux density measurement in the fit, as we lack single dish continuum data (we only have continuum data from the 12 m \& 7 m arrays). Therefore, we do not recover all the flux from the region in these ALMA continuum observations, which leads to {a} lower overall integrated flux density measurement. 


In the UV/optical/NIR bands (i.e.\ $\lambda \leq8\mu m$), the emission from the IRAS 19132+1035 region is dominated by compact point sources, the brightest of which coincides with the peak of the radio continuum emission. Therefore, emission in these bands appears to be mainly stellar emission.

\subsection{CO line emission from IRAS 19132+1035}
\label{sec:CO}

Integrated intensity maps of the $^{12}$CO, $^{13}$CO, and C$^{18}$O emission across the IRAS 19132+1035 region are shown in the bottom panels of Figure~\ref{fig:grsconts}. In these maps, we detect bright emission from the different isotopologues of CO, coincident with the radio continuum feature. However, we see very little emission to the south of the radio continuum feature (other than in $^{12}$CO). 

\renewcommand\tabcolsep{4pt}
 \begin{table}
\caption{CO Line Emission Properties from Gaussian Fits}\quad
\centering
\begin{tabular}{ cccccc }
 \hline\hline
 {\bf Reg.}&{\bf Line}&{\bf \# of}&{\bf $T_{\rm \bf p}$$^b$}&{\bf FWHM}&{\bf $\mathbf{V_{\rm \bf c}}$$^c$}\\
 &&{\bf Comp.$^a$}&{\bf (K)}&{($\mathbf{{\rm \bf km\,s}^{-1}}$)}&{($\mathbf{{\rm \bf km\,s}^{-1}}$)}\\[0.15cm]
 \hline
A&$^{12}$CO&2&$10.60^{+0.05}_{-0.05}$&$2.76^{+0.03}_{-0.03}$&$65.73^{+0.02}_{-0.01}$\\[0.15cm]
&&&$16.23^{+0.06}_{-0.06}$&$2.68^{+0.01}_{-0.01}$&$68.54^{+0.01}_{-0.01}$\\[0.15cm]
&$^{13}$CO&2&$10.31^{+0.10}_{-0.10}$&$1.85^{+0.02}_{-0.02}$&$66.32^{+0.01}_{-0.01}$\\[0.15cm]
&&&\phantom{0}$7.45^{+0.07}_{-0.07}$&$2.10^{+0.03}_{-0.03}$&$68.27^{+0.02}_{-0.02}$\\[0.15cm]
&C$^{18}$O&2&\phantom{0}$2.05^{+0.05}_{-0.05}$&$0.90^{+0.03}_{-0.03}$&$66.04^{+0.01}_{-0.01}$\\[0.15cm]
&&&\phantom{0}$3.75^{+0.02}_{-0.02}$&$2.10^{+0.03}_{-0.03}$&$67.59^{+0.01}_{-0.01}$\\[0.15cm]
\\[-0.15cm]
B&$^{12}$CO&2&$10.16^{+0.02}_{-0.02}$&$2.73^{+0.01}_{-0.01}$&$65.54^{+0.01}_{-0.01}$\\[0.15cm]
&&&$19.75^{+0.03}_{-0.02}$&$2.43^{+0.01}_{-0.01}$&$68.43^{+0.003}_{-0.003}$\\[0.15cm]
&$^{13}$CO&2&\phantom{0}$9.56^{+0.06}_{-0.07}$&$2.77^{+0.03}_{-0.03}$&$66.91^{+0.02}_{-0.02}$\\[0.15cm]
&&&\phantom{0}$3.49^{+0.2}_{-0.2}$&$1.63^{+0.04}_{-0.05}$&$68.61^{+0.02}_{-0.02}$\\[0.15cm]
&C$^{18}$O&2&\phantom{0}$1.13^{+0.16}_{-0.11}$&$1.32^{+0.10}_{-0.10}$&$66.22^{+0.07}_{-0.04}$\\[0.15cm]
&&&\phantom{0}$3.21^{+0.05}_{-0.07}$&$1.98^{+0.05}_{-0.06}$&$67.61^{+0.05}_{-0.03}$\\[0.15cm]
\\[-0.15cm]
C&$^{12}$CO&1&$22.10^{+0.02}_{-0.03}$&$3.70^{+0.004}_{-0.004}$&$67.37^{+0.002}_{-0.002}$\\[0.15cm]
&$^{13}$CO&1&$15.25^{+0.01}_{-0.01}$&$2.83^{+0.002}_{-0.002}$&$66.91^{+0.001}_{-0.001}$\\[0.15cm]
&C$^{18}$O&1&\phantom{0}$5.25^{+0.01}_{-0.01}$&$1.88^{+0.003}_{-0.003}$&$67.07^{+0.001}_{-0.001}$\\[0.15cm]
\\[-0.15cm]
D&$^{12}$CO&1&$11.05^{+0.009}_{-0.007}$&$2.89^{+0.002}_{-0.003}$&$67.40^{+0.001}_{-0.001}$\\[0.15cm]
&$^{13}$CO&1&$3.02^{+0.007}_{-0.006}$&$2.03^{+0.005}_{-0.004}$&$67.56^{+0.002}_{-0.002}$\\[0.15cm]
&C$^{18}$O&1&$0.25^{+0.009}_{-0.009}$&\phantom{0}$2.04^{+0.08}_{-0.07}$&\phantom{0}$67.72^{+0.04}_{-0.04}$\\[0.15cm]
\hline
\end{tabular}\\
\begin{flushleft}
{$^a$Number of Gaussian components needed to fit the line.}\\
{$^b$Peak intensity of Gaussian components.}\\
{$^c$Central velocity of Gaussian components.}
\end{flushleft}
\label{table:Co_line_vals}
\end{table}
\renewcommand\tabcolsep{6pt}

As the CO emission tracks where most of the molecular gas is located, we expect signatures of a jet-ISM interaction 
to appear best within the CO emission morphology.
In particular, in the framework of the jet-ISM interaction model presented in \cite{kai04}, the radio continuum feature is housed inside a jet-blown bubble in the molecular cloud, and the entire IRAS source is thought to represent shock-heated ISM material located near the jet impact zone. The key morphological features we then anticipate to observe are a cavity structure surrounding the linear non-thermal emission feature (possibly extending towards the peak of the radio continuum), a ring-like bow shock feature hugging the sharp southern edge of the entire radio continuum, and shocked molecular gas to the south of the radio continuum feature (see Figure~\ref{fig:grs} \textit{right}). However, we see none of these expected features within the CO emission. While this does not rule out a jet-ISM interaction in this region, it strongly suggests that the entire IRAS 19132+1035 region (or at least the flat-spectrum radio continuum feature) is not predominantly shaped by such an interaction.

To investigate the possibility of a jet-ISM interaction on smaller scales (i.e.\ smaller than the extent of the radio continuum feature), we examined the CO emission properties, namely spectra and line ratios, across different regions of interest in the imaged field; the base of the non-thermal jet feature (A), the radio peak of the non-thermal jet-feature (B), the peak of the radio continuum (C), and an off position well away from the radio continuum feature (D; shown in Figure~\ref{fig:CO_mom0_spec} \textit{left}). To characterize the spectral line properties, we fit each detected line with one or more Gaussian components, where we have estimated the uncertainties on the spectral data points for each line by taking the median absolute deviation of spectral data points well away from the line emission (see Table~\ref{table:Co_line_vals}).

The line profiles and intensities of the CO emission clearly vary across the IRAS 19132+1035 region (see Figure~\ref{fig:CO_mom0_spec} \textit{right}).
Regions A and B display CO lines with double peaked, more asymmetric profiles, when compared to regions C and D. The peak intensities also differ between regions, where the brightest emission from all the isotopologues occurs in region C, coincident with the peak of the radio continuum emission.

In a previous molecular line study of this region conducted with the IRAM 30m telescope (which had lower sensitivity, as well as lower spectral and angular resolution, when compared to our ALMA observations), \cite{chat01} detected the $^{12}$CO ($J=2-1$) and $^{13}$CO ($J=2-1$) transitions at multiple positions along a slice through the radio jet feature.
Our new spectral measurements of the CO emission in the IRAS 19132+1035 region are consistent with those reported in this previous study, in terms of the central velocities and peak intensities of the $^{12}$CO and $^{13}$CO lines ($\sim67~{\rm km\,s}^{-1}$ and $\sim$ 20, 11 K, respectively), as well as the asymmetric line profiles observed near the jet feature.

\begin{figure}
\begin{center}
\includegraphics[width=1\columnwidth]{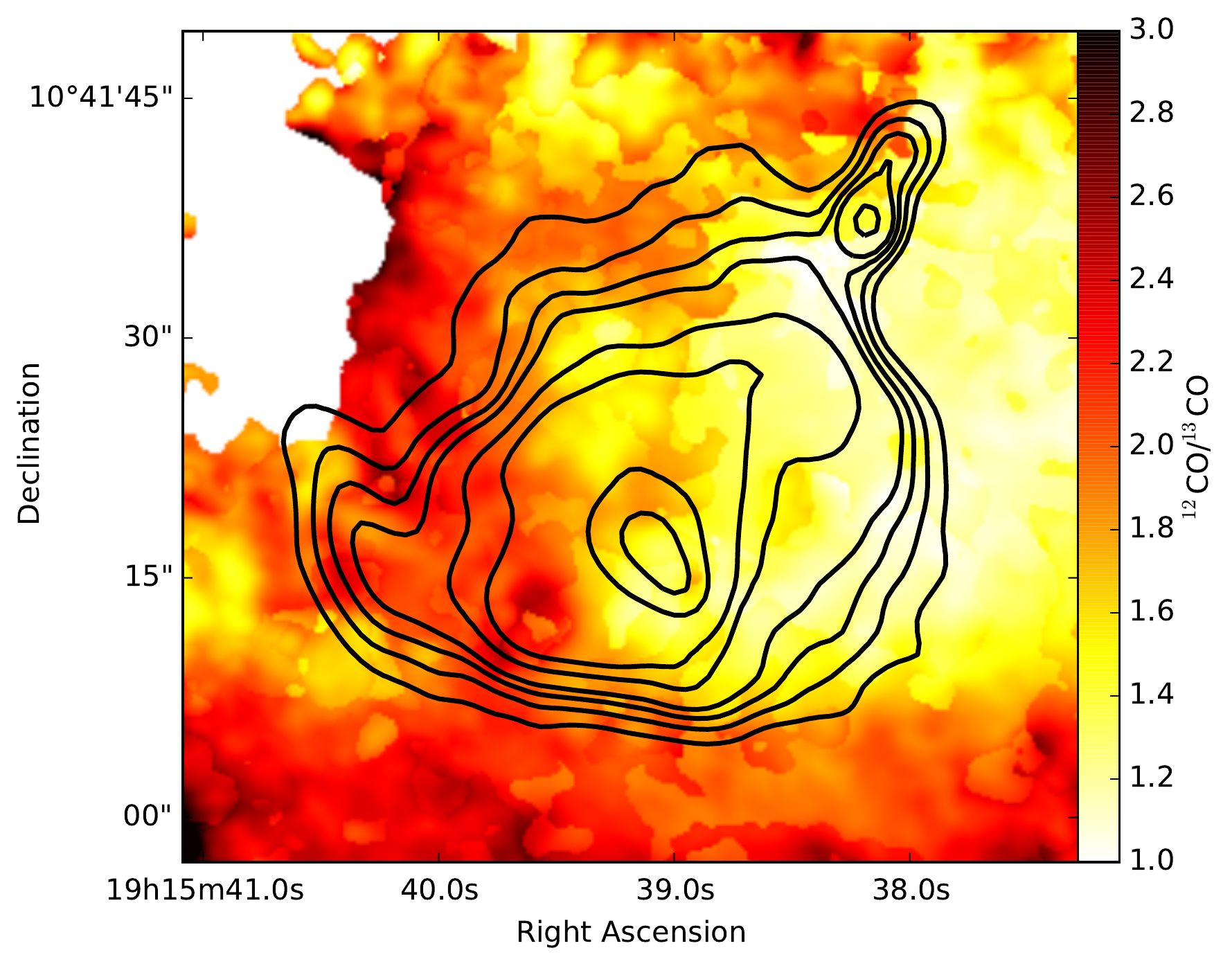}
\caption{\small \label{fig:CO_ratios}  $^{12}$CO/$^{13}$CO isotopologue ratio map of the IRAS 19132$+$1035 region. The colour bar indicates the line ratio values. The contours represent the VLA radio frequency contours; levels of 0.2, 0.3, 0.4, 0.5, 0.75, 1, 2, and 3 ${\rm mJy\,bm}^{-1}$ (see Figure~\ref{fig:grsconts}). Regions where the $^{13}$CO emission is $<5$ K are masked. 
The region located at the base of the radio jet feature displays an atypical line ratio, $^{12}$CO/$^{13}$CO$\sim 1$, signifying very optically thick (dense) gas in this zone.}
\end{center}
\end{figure}

The CO line ratios also vary substantially across the IRAS 19132+1035 region (see Figure~\ref{fig:CO_ratios}\footnote{To create this ratio map, we use the $^{13}$CO as a template to find the velocity channel with the maximum intensity in each pixel, then compute the line ratio for each pixel using the intensity in the matching velocity channel of the $^{12}$CO spectrum. This procedure ensures that we are not comparing emission at different velocities.}). In particular, region A displays an atypical ratio of $^{12}$CO/$^{13}$CO$\sim 1$. This line ratio indicates the presence of optically thick (likely very dense) gas, preferentially located in the region coincident with the base of the non-thermal jet feature.

\begin{figure*}
\begin{center}
\includegraphics[width=2\columnwidth]{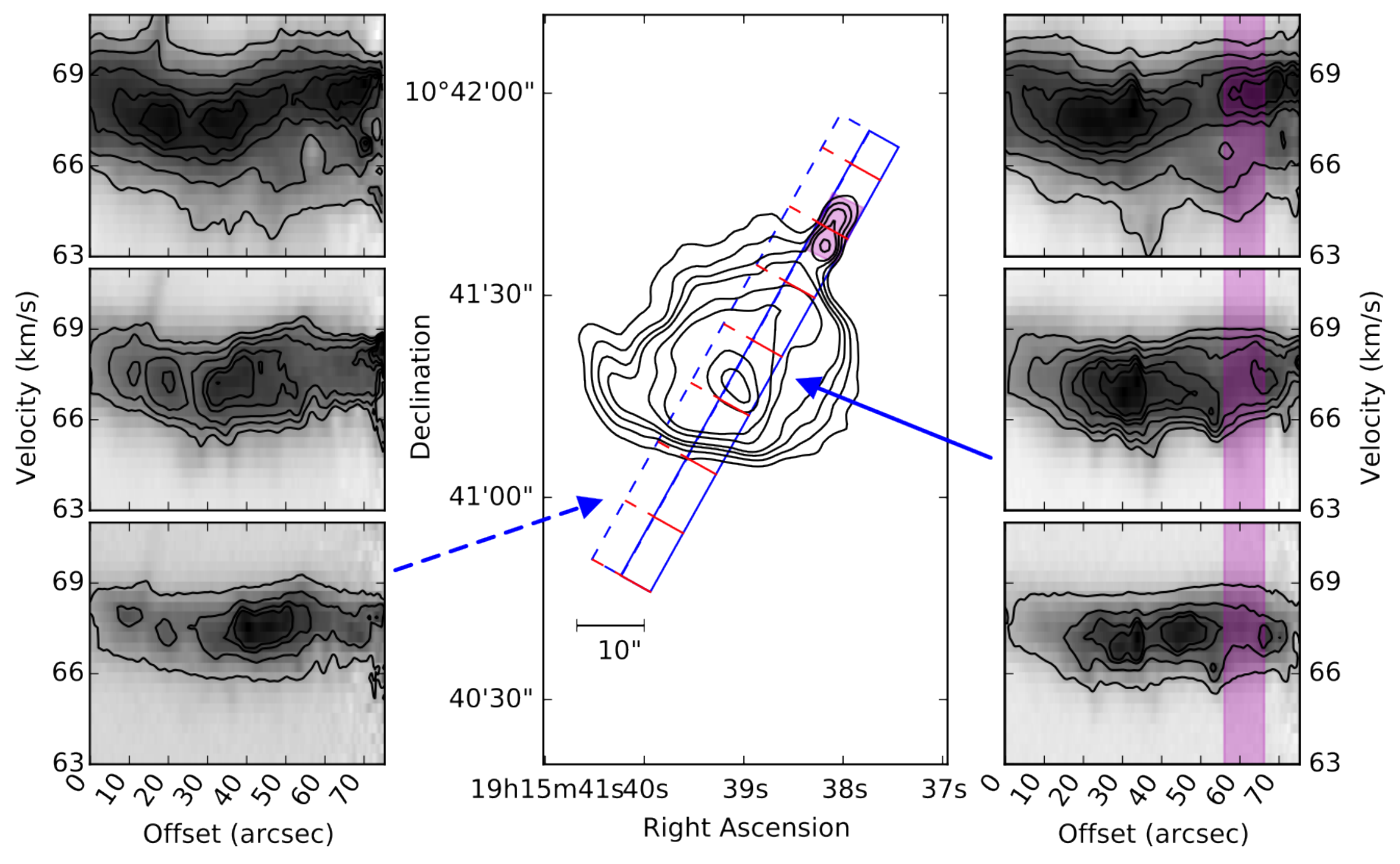}
\caption{\small \label{fig:CO_PV}  Kinematic analysis of the CO emission in the IRAS 19132$+$1035 region, over a slice through the radio jet feature (solid blue rectangle) and a slice offset from the radio jet feature (dotted blue rectangle). The {\it left} and {\it right} panels display the PV diagrams of $^{12}$CO ({\it top}), $^{13}$CO ({\it middle}), and C$^{18}$O ({\it bottom}) from the respective slices. Contours are  5.0, 10.0, 15.0, 17.5, and 20.0 K for $^{12}$CO, 5.0, 6.5, 8.0, 10.0, 11.0, and 12.5 K for $^{13}$CO, and 1.0, 2.5, 3.5, and 4.0 K for C$^{18}$O in both the {\it left and right panels}.
The {\it middle panel} displays the location of the slices through the CO cubes, where the contours represent the VLA radio frequency contours; levels of 0.2, 0.3, 0.4, 0.5, 0.75, 1, 2, and 3 ${\rm mJy\,bm}^{-1}$ (see Figure~\ref{fig:grsconts}). The purple shading in the {\it right and middle panels} indicate the spatial location of the radio jet feature. The red ticks in the middle panel correspond to the offset along the slices, in increments of 10\arcsec (from the lower left; 0, 10, 20, 30, 40, 50, 60, 70\arcsec). 
Kinematic features unique to the radio jet feature region (i.e.\ a distinct hole in the $^{13}$CO and C$^{18}$O emission, and an elongated lobe extended towards lower velocities), suggest that the gas in this region is being punched from behind (the direction of the BHXB jet), resulting in the excavation of a jet-blown cavity. }
\end{center}
\end{figure*}

To examine the kinematics of the molecular gas near our regions of interest, we created Position-Velocity (PV) diagrams of $^{12}$CO, $^{13}$CO, and C$^{18}$O, from a slice through the linear non-thermal jet feature (see Figure~\ref{fig:CO_PV} \textit{right}). Within these diagrams, we observe a distinct hole in the $^{13}$CO and C$^{18}$O molecular emission, spatially coincident with the location of the non-thermal jet feature. Additionally, we observe an elongated lobe, extended towards lower velocities, located at the inner edge of the jet feature. These PV features are unique to the jet feature region (i.e.\ they do not appear in the shifted PV slices; see Figure~\ref{fig:CO_PV} \textit{left}), and indicate not only that the gas is being spatially displaced away from the linear jet feature region, but also that this displaced gas is being pushed to lower velocities. Both of these kinematic properties are consistent with the gas in the jet feature region being punched from behind (i.e.\ in line with the direction of the BHXB jet), resulting in the excavation of a potential jet-blown cavity. 
We note that these enhancements/depletions in the CO emission are not seen as clearly in the $^{12}$CO compared to the $^{13}$CO and C$^{18}$O emission. This is likely a result of opacity effects; the $^{12}$CO emission is optically thick, and this will in turn make it difficult to detect such enhancements/depletions in the $^{12}$CO emission.

\subsection{Shock-tracing line emission from IRAS 19132+1035}
\label{sec:st}
\begin{figure*}
\begin{center}
 \includegraphics[width=2\columnwidth]{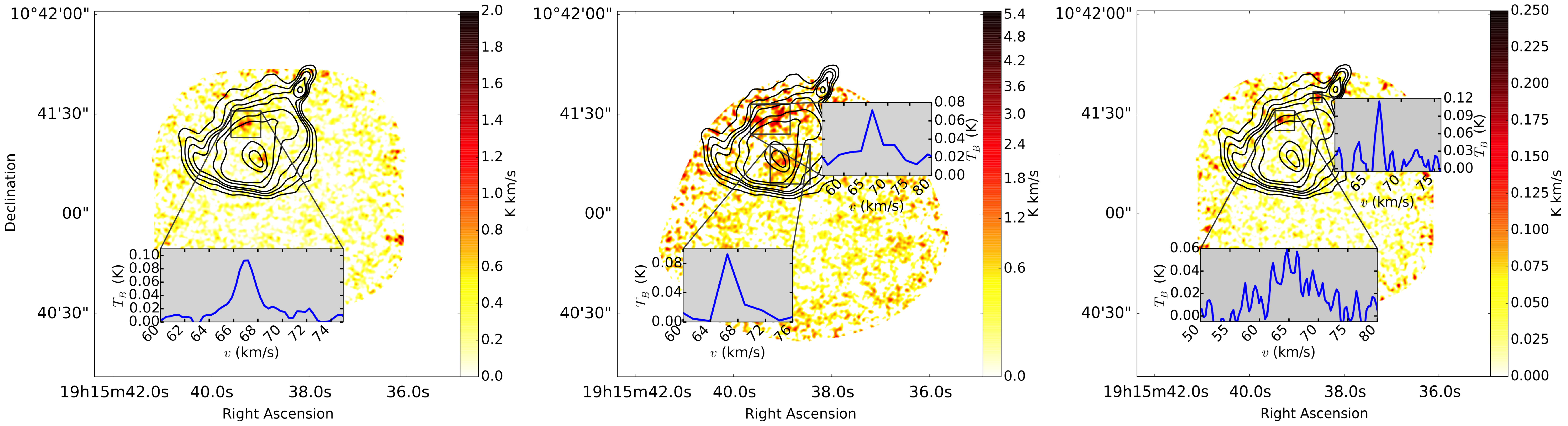}
 \caption{\small \label{fig:st_mom0} Emission from the shock-tracing molecules detected in the IRAS 19132$+$1035 region. The panels from {\it left to right} display the CH$_3$OH ($J=4_{2,2}-3_{1,2}$),  CS ($\nu=0$, $J=5-4$), and SiO ($\nu=0$, $J=5-4$), integrated intensity maps. The contours in all the panels are the VLA radio frequency contours; levels of 0.2, 0.3, 0.4, 0.5, 0.75, 1, 2, and 3 ${\rm mJy\,bm}^{-1}$ (see Figure~\ref{fig:grsconts}).
 The colour bars indicate the integrated intensity in units of ${\rm K\,km\,s}^{-1}$.
 Inset panels show spectra of each line, in regions where we detect significant emission (marked by the black boxes). Note that the CS line has a noticeably lower spectral resolution, as this line was detected within the continuum spectral window. Due to the compact nature of the northernmost SiO detection, we also provide an alternate version of the SiO integrated intensity map zoomed in on this feature in Figure~\ref{fig:zoomsio}. We observe shock-tracing emission with varying line widths and intensities across IRAS 19132+1035, suggesting multiple feedback sources are powering the region.
 }
 \end{center}
 \end{figure*}

Integrated intensity maps and spectra for the CH$_3$OH, CS, and SiO transitions detected in the IRAS 19132+1035 region are shown in Figure~\ref{fig:st_mom0}, with the spectral line characteristics from Gaussian fits displayed in Table~\ref{table:st_line_vals}. We did not detect the N$_2$D$^+$ molecule in the region, where we place an estimated $3\sigma$ upper limit on the integrated intensity across the IRAS 19132+1035 region of $0.69\, {\rm K\,km\,s}^{-1}$.

The majority of the emission, showing the brightest integrated intensities, from the CH$_3$OH, CS, and SiO molecules are confined to a region located to the north of the radio continuum peak, coincident with sub-mm continuum emission. Additionally, the CS line is detected in another region near the radio continuum peak, and the SiO line is detected in a compact region near the base of the jet feature.
These shock tracing emission lines have much lower peak intensities than the CO lines; the CH$_3$OH line has a peak intensity of $\sim80$ mK, CS has a peak intensity between $\sim50$--$120$ mK, and SiO has a peak intensity of $\sim40$--$110$ mK.

\begin{figure}
\begin{center}
\includegraphics[width=1\columnwidth]{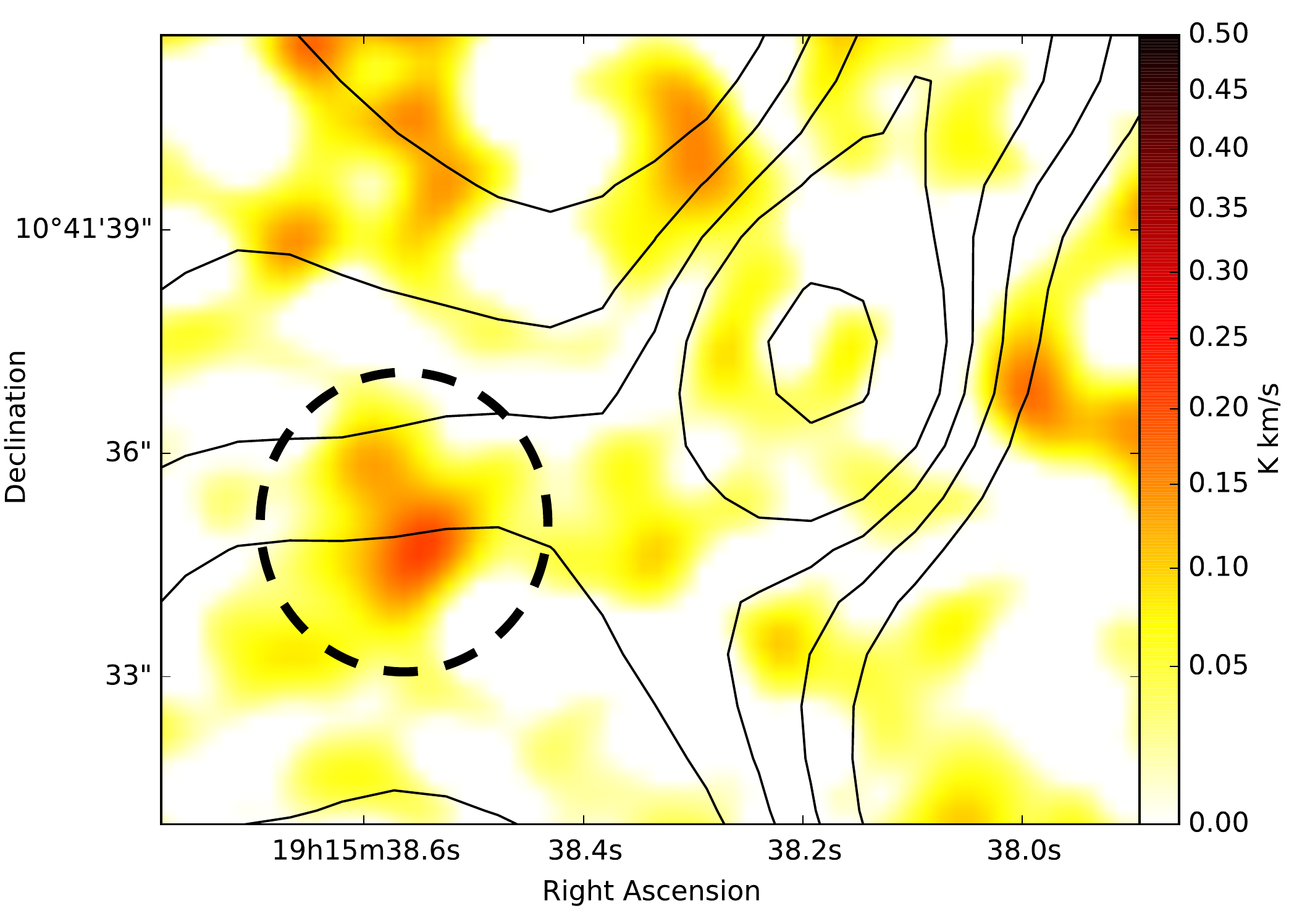}
\caption{\small \label{fig:zoomsio} Integrated intensity map of SiO ($\nu=0$, $J=5-4$) emission in IRAS 19132+1035, zoomed in to show the northernmost detection (indicated by the black circle). The integrated intensity map covering the entire IRAS 19132+1035 region can be seen in the rightmost panel of Figure~\ref{fig:st_mom0}. We are confident that this detection is real and not simply a noise feature; the emission spans multiple channels, the peak intensity of the line is $>5\sigma$ above the estimated noise level ($\sim0.02$ K), and the central velocity matches that of our other shock tracing line detections.
}
\end{center}
\end{figure}

While the CH$_3$OH and CS lines appear to have similar line widths to that of the CO in region C
($\sim2$--$3\,{\rm km\,s}^{-1}$), the SiO detections show both significantly wider (southern detection) and narrower (northern detection\footnote{Despite the narrow line width and low integrated intensity for the northern SiO detection (when compared to the southern SiO detection), we are confident this detection is real, as the central velocity matches that of the other shock tracing line detections, and the peak intensity of the line is $>5\sigma$ above the estimated noise level ($\sim0.02$ K) in the spectrum.}; also see Figure~\ref{fig:zoomsio}) line widths when compared to CO ($\sim10\,{\rm km\,s}^{-1}$ and $\sim1\,{\rm km\,s}^{-1}$).

 \renewcommand\tabcolsep{4pt}
 \begin{table}
\caption{Shock Tracing Line Emission Properties from Gaussian Fits}\quad
\centering
\begin{tabular}{ lcccc }
 \hline\hline
 {\bf Line}&{\bf \# of}&{\bf $T_{\rm \bf p}$$^b$}&{\bf FWHM}&{\bf $\mathbf{V_{\rm \bf c}}$$^c$}\\
 &{\bf Comp.$^a$}&{\bf (K)}&{($\mathbf{{\rm \bf km\,s}^{-1}}$)}&{($\mathbf{{\rm \bf km\,s}^{-1}}$)}\\[0.15cm]
 \hline
CH$_3$OH&1&$0.08^{+0.005}_{-0.005}$&\phn$2.63^{+0.19}_{-0.17}$&$67.00^{+0.07}_{-0.05}$\\[0.15cm]
SiO North&1&$0.11^{+0.02}_{-0.02}$&\phn$1.01^{+0.24}_{-0.18}$&$66.66^{+0.10}_{-0.09}$\\[0.15cm]
SiO South&1&$0.04^{+0.003}_{-0.003}$&$10.01^{+0.88}_{-0.86}$&$65.56^{+0.19}_{-0.21}$\\[0.15cm]
CS North$^d$&1&$0.05^{+0.03}_{-0.01}$&\phn$2.70^{+0.67}_{-0.94}$&$66.73^{+0.47}_{-0.57}$\\[0.15cm]
CS South&1&$0.12^{+0.10}_{-0.02}$&\phn$2.12^{+0.83}_{-0.83}$&$67.03^{+0.38}_{-0.59}$\\[0.15cm]
\hline
\end{tabular}\\
\begin{flushleft}
{$^a$Number of Gaussian components needed to fit the line.}\\
{$^b$Peak intensity of Gaussian components.}\\
{$^c$Central velocity of Gaussian components.}\\
{$^d$Note that an additional constant baseline flux component was required when fitting this line; its amplitude was $0.022^{+0.002}_{-0.002}$ K.}
\end{flushleft}
\label{table:st_line_vals}
\end{table}
\renewcommand\tabcolsep{6pt}

In their previous molecular line study of IRAS 19132+1035, \cite{chat01} reported the detection of the shock tracing molecules, CS ($J=2-1$) and SiO ($J=2-1$ \& $J=3-2$), at locations near the radio continuum peak and the non-thermal jet feature, respectively.
The location, central velocity ($\sim67\,{\rm km\,s}^{-1}$) and line width ($\sim2-3\,{\rm km\,s}^{-1}$) of our southernmost detection of CS ($J=5-4$), is consistent with the previous detection of the $J=2-1$ transition of this molecule. Similarly, the spectral line properties of our northernmost detection of SiO ($J=5-4$) also appears to be consistent with the previous detection of the $J=2-1$ and $J=3-2$ transitions by \cite{chat01}, especially in terms of the observed narrow line widths of only a few km\,s$^{-1}$.

Given its proximity to the non-thermal jet feature (i.e.\ the suspected jet impact site), and the compact nature of the emission region, our northernmost SiO detection may originate in a shock produced as the BHXB jet collides with the ISM. However, the narrow line width implies the shock velocity must be quite low\footnote{See \S\ref{sec:compare} for a discussion on this low shock velocity.}. 
The remainder of the shock tracing emission we detect is only found in regions located well away from the non-thermal jet feature. Therefore, these detections are unlikely to be associated with a potential jet-ISM interaction site in IRAS 19132+1035 and must originate from another source of feedback in the region, likely from high mass star formation (see \S\ref{sec:hmsf}). This conclusion is also supported by the vastly differing line widths between the SiO detections in these two shocked emission regions. The molecular gas to the south-east of the jet feature likely has a much faster shock compared to the molecular gas near the base of the jet feature (see Figure~\ref{fig:st_mom0} and Table~\ref{table:st_line_vals}), suggesting that these shock features are powered by feedback from different objects.

  \begin{figure*}
\begin{center}
 \includegraphics[width=2\columnwidth]{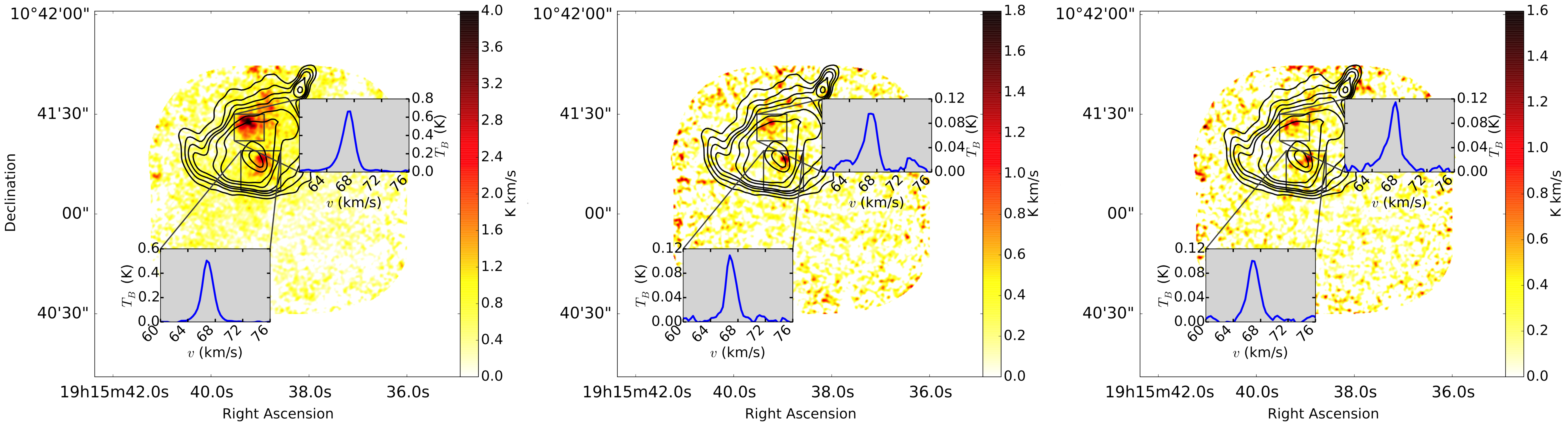}
 \caption{\small \label{fig:H2CO_mom0} H$_2$CO emission from the IRAS 19132$+$1035 region. The panels from {\it left to right} display the ${\rm H}_{2}{\rm CO}$ ($J=3_{0,3}-2_{0,2}$), ${\rm H}_{2}{\rm CO}$ ($J=3_{2,1}-2_{2,0}$), and ${\rm H}_{2}{\rm CO}$ ($J=3_{2,2}-2_{2,1}$), integrated intensity maps. The contours in all the panels are the VLA radio frequency contours; levels of 0.2, 0.3, 0.4, 0.5, 0.75, 1, 2, and 3 ${\rm mJy\,bm}^{-1}$ (see Figure~\ref{fig:grsconts}). The colour bars indicate the integrated intensity in units of ${\rm K\,km\,s}^{-1}$. Inset panels show spectra of each line in the brightest emission zones (marked by the black boxes).
 A lack of ${\rm H}_{2}{\rm CO}$ emission within, and at the base of, the non-thermal jet feature indicates the presence of preferentially colder molecular gas, when compared to the surrounding zones.
 }
 \end{center}
 \end{figure*}

\subsection{${\rm H}_{2}{\rm CO}$ line emission from IRAS 19132+1035}
\label{sec:h2co}

Integrated intensity maps and spectra for the three ${\rm H}_{2}{\rm CO}$ lines detected in the IRAS 19132+1035 region are shown in Figure~\ref{fig:H2CO_mom0}, with the spectral line characteristics from Gaussian fits displayed in Table~\ref{table:h2co_line_vals}.

The majority of the emission from ${\rm H}_{2}{\rm CO}$ is confined to regions coincident with the sub-mm continuum emission (i.e.\ centred on the radio continuum peak, and a region to the north of the radio peak). The brightest ${\rm H}_{2}{\rm CO}$ line displays a peak intensity of $\sim 0.6$ K, while the weaker lines display peak intensities of $\sim 0.1$ K. The line widths of all three transitions are similar to those of the CH$_3$OH and CS molecules present in the same region ($\sim2\,{\rm km\,s}^{-1}$). 
\renewcommand\tabcolsep{2pt}
 \begin{table}
\caption{H$_2$CO Line Emission Properties from Gaussian Fits}\quad
\centering
\begin{tabular}{ ccccc }
 \hline\hline
 {\bf H$_2$CO}&{\bf \# of}&{\bf $T_{\rm \bf p}$$^b$}&{\bf FWHM}&{\bf $\mathbf{V_{\rm \bf c}}$$^c$}\\
 {\bf Transition}&{\bf Comp.$^a$}&{\bf (K)}&{($\mathbf{{\rm \bf km\,s}^{-1}}$)}&{($\mathbf{{\rm \bf km\,s}^{-1}}$)}\\[0.15cm]
 \hline
 $J=3_{0,3}-2_{0,2}$ North&1&$0.64^{+0.003}_{-0.003}$&$2.22^{+0.01}_{-0.01}$&$67.15^{+0.006}_{-0.005}$\\[0.15cm]
$J=3_{0,3}-2_{0,2}$ South&1&$0.49^{+0.003}_{-0.004}$&$2.00^{+0.02}_{-0.02}$&$66.89^{+0.007}_{-0.006}$\\[0.15cm]
$J=3_{2,1}-2_{2,0}$ North&1&$0.10^{+0.004}_{-0.004}$&$2.36^{+0.14}_{-0.14}$&$67.07^{+0.05}_{-0.05}$\\[0.15cm]
$J=3_{2,1}-2_{2,0}$ South&1&$0.11^{+0.003}_{-0.002}$&$1.86^{+0.06}_{-0.06}$&$66.98^{+0.02}_{-0.02}$\\[0.15cm]
$J=3_{2,2}-2_{2,1}$ North&1&$0.10^{+0.005}_{-0.005}$&$2.37^{+0.15}_{-0.14}$&$67.20^{+0.05}_{-0.04}$\\[0.15cm]
$J=3_{2,2}-2_{2,1}$ South&1&$0.09^{+0.003}_{-0.002}$&$2.18^{+0.06}_{-0.07}$&$66.96^{+0.03}_{-0.02}$\\[0.15cm]
\hline
\end{tabular}\\
\begin{flushleft}
{$^a$Number of Gaussian components needed to fit the line.}\\
{$^b$Peak intensity of Gaussian components.}\\
{$^c$Central velocity of Gaussian components.}\\
\end{flushleft}
\label{table:h2co_line_vals}
\end{table}
\renewcommand\tabcolsep{6pt}

To derive an estimate of the physical properties of the molecular gas (i.e.\ temperature and density) in the IRAS 19132+1035 region, we model the ${\rm H}_{2}{\rm CO}$ lines with RADEX radiative transfer models, using the \textsc{pyspeckit} python package\footnote{https://github.com/pyspeckit/pyspeckit}.
This  package implements RADEX model grids over a range of densities, column densities, and temperatures\footnote{In our modelling, we implement pre-computed RADEX model grids over the following ranges for temperature, density, and column density, $5<T<205$ K, $10^{2}<n<10^{7}{\rm cm}^{-3}$, and $10^{10}<N<10^{17}\, {\rm cm}^{-2}$.} to fit a five parameter model to the ${\rm H}_{2}{\rm CO}$ spectrum: line intensity, line width, temperature, density, and column density.
For this modelling, we simultaneously fit the ${\rm H}_{2}{\rm CO}$($J=3_{0,3}-2_{0,2}$), ${\rm H}_{2}{\rm CO}$ ($J=3_{2,2}-2_{2,1}$), and ${\rm H}_{2}{\rm CO}$ ($J=3_{2,1}-2_{2,0}$) lines in regions where we detect significant ${\rm H}_{2}{\rm CO}$ emission. We define a significant detection of ${\rm H}_{2}{\rm CO}$ as regions where emission in the $T_{\rm max}$ map of the brightest transition ($J=3_{0,3}-2_{0,2}$) reaches a $\geq 3\sigma$ level. This translates to regions where the ${\rm H}_{2}{\rm CO}$ ($J=3_{0,3}-2_{0,2}$) emission is $>0.25$ K.

\begin{figure}
\begin{center}
\includegraphics[width=1\columnwidth]{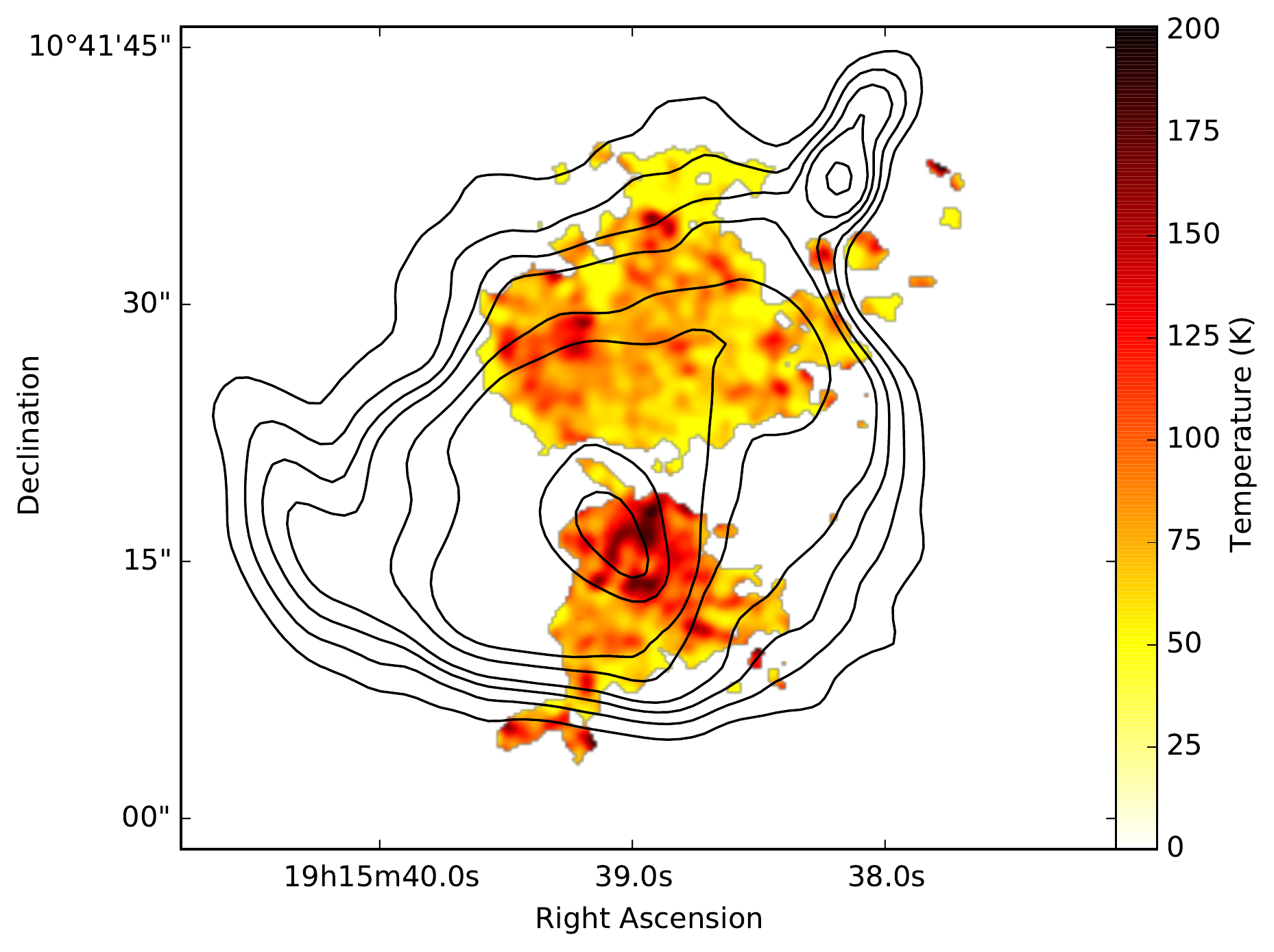}
\caption{\small \label{fig:radex} Temperature map of the molecular gas in the IRAS 19132$+$1035 region, obtained from fitting the H$_2$CO lines with RADEX models. In this fit, we fix the density in the region to $n=10^4\mbox{ cm}^{-3}$. The temperature map shown here is masked to regions where we detected H$_2$CO emission (see \S\ref{sec:h2co} for details). The contours are the VLA radio frequency contours; levels of 0.2, 0.3, 0.4, 0.5, 0.75, 1, 2, and 3 ${\rm mJy\,bm}^{-1}$ (see Figure~\ref{fig:grsconts}). The colour bar indicates the temperature in units of K. The largest temperatures in the region appear to coincide with the peak of the radio continuum. 
}
\end{center}
\end{figure}

When performing the RADEX modelling, with all of the parameters left free, we are unable to accurately map the temperature in the region, as variations in density appear to be driving artificial variations in the fitted temperature. This result is likely due to the degeneracies between the temperature and density parameters in the model. Therefore, we choose to fix the density at the central value of $n=10^{4}\,{\rm cm}^{-3}$, and refit the spectrum. A temperature map of the region output from this process is shown in Figure~\ref{fig:radex}. Based on this map, the hottest molecular gas in the region appears to be clustered around the radio continuum peak. Additionally, there is a distinct lack of ${\rm H}_{2}{\rm CO}$ emission within, and at the base of, the non-thermal jet feature (see Figure~\ref{fig:H2CO_mom0}). {This indicates the absence of hot molecular gas within and at the base of the jet feature, when compared to the surrounding zones.} 

\subsection{Column densities}
\label{sec:column}

To accurately map the column density of the molecular gas in the IRAS 19132+1035 region, we use the ${\rm C}^{18}{\rm O}$ line emission.
A detailed derivation of this column density is provided in Appendix \ref{sec:column_app}.
Figure~\ref{fig:colco} shows the C$^{18}$O column density map of the IRAS 19132+1035 region. This column density should give a good indication of where the majority of the gas mass is located in the region. Similar to what is observed in the CO integrated intensity maps, most of the molecular gas coincides with the radio continuum feature. Additionally, the gas with highest column densities (indicated by the green contours) appears to surround and hug the linear jet feature, consistent with the gas building up along the edges of a potential jet-blown cavity in that region.

We opt to calculate the H$_2$ surface density from the molecular emission, since the dust maps lack sufficient resolution to map the small-scale structure of the region.  Assuming a fractional abundance of $\mathrm{{}^{12}CO/H_2} = 2\times 10^{-4}$ and $\mathrm{^{16}O/{}^{18}O}=300$ for $R_{\mathrm{gal}}=6~\mathrm{kpc}$ \citep{wilsonrood}, yields a ratio of $\mathrm{H_2/C^{18}O} = 1.5\times 10^{6}$.  The implied H$_2$ column densities of the molecular gas can be obtained by scaling the values in Figure~\ref{fig:colco} by this factor.

\begin{figure}
\begin{center}
\includegraphics[width=1\columnwidth]{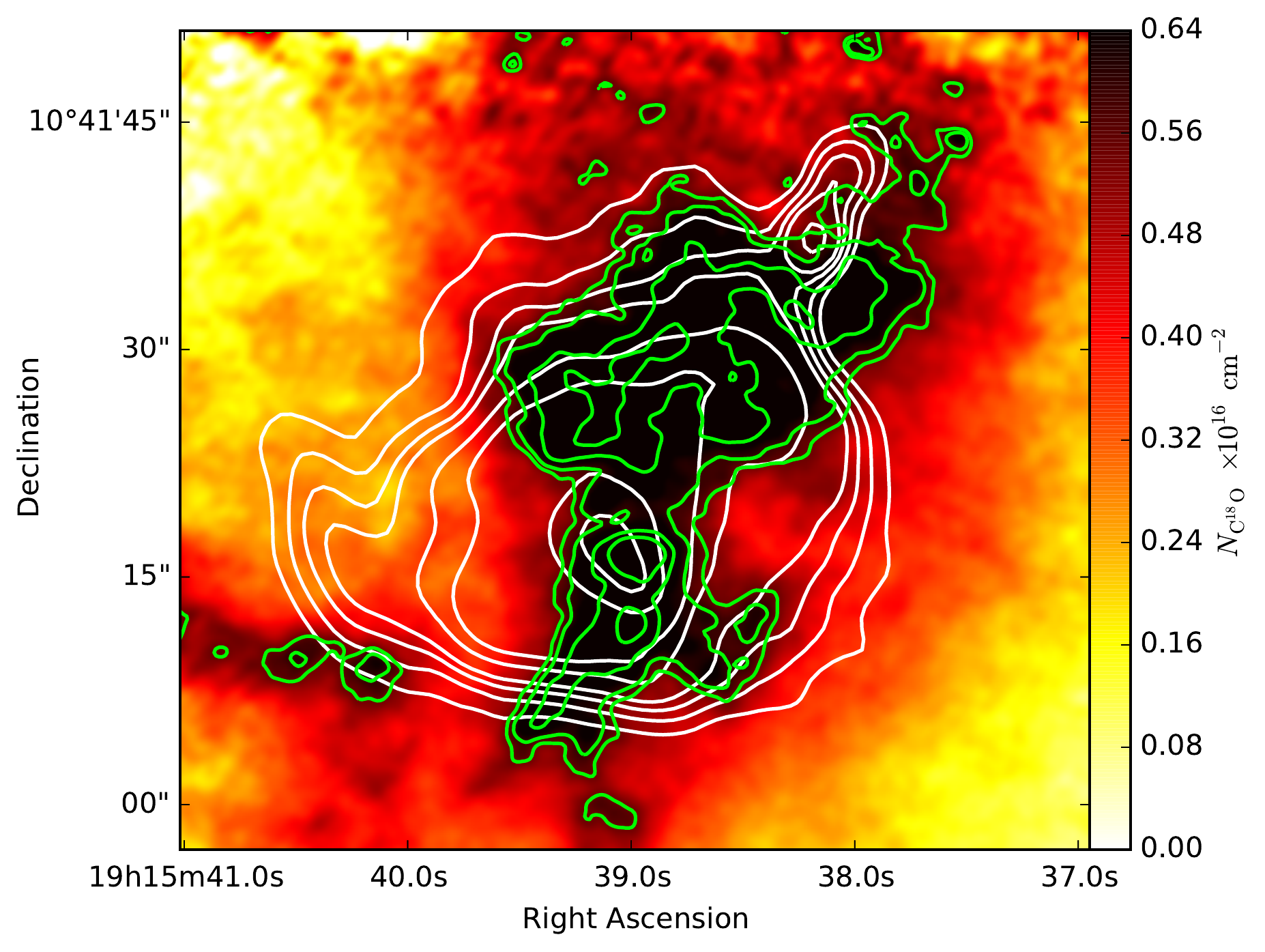}
\caption{\small \label{fig:colco} C$^{18}$O column density map of the IRAS 19132+1035 region. The colour bar represents the C$^{18}$O column density in units of $10^{16}\,{\rm cm}^{-2}$. The green contours indicate regions with the highest column densities, at levels of 5.5, 6.0, 7.0, and 8.0 $\times10^{15}\,{\rm cm}^{-2}$. The white contours represent the VLA radio continuum contours; levels of 0.2, 0.3, 0.4, 0.5, 0.75, 1, 2, and 3 ${\rm mJy\,bm}^{-1}$ (see Figure~\ref{fig:grsconts}). These data should represent the column densities of H$_2$ after scaling the map values by a factor of $1.5\times 10^{6}$. Molecular gas with the highest column densities appears to hug the jet feature, consistent with the presence of a jet-blown cavity in this region.}
\end{center}
\end{figure}

\section{Discussion} 
\label{sec:discuss}
\subsection{Evidence for a jet-ISM interaction}
In this work, we have presented extensive new data tracing the molecular line emission and mapping the ISM conditions across the IRAS 19132+1035 region. These data provide multiple lines of new evidence supporting an association between the IRAS 19132+1035 region and the GRS 1915+105 jet. In particular, our key new evidence for a jet-ISM interaction in the IRAS 19132+1035 zone are as follows:
\begin{enumerate}
\item CO kinematics indicate  the gas in the region covered by the radio jet feature is being hit from behind (in line with the direction of the GRS 1915+105 jet), resulting in the creation of a jet-blown cavity in the molecular gas at this location (see \S\ref{sec:CO} and Figure~\ref{fig:CO_PV}).
\item CO line ratios indicate the presence of over-dense gas at the apex of the suspected jet impact zone (i.e.\ the base of the radio jet feature), consistent with gas being excavated to form a cavity in the radio jet feature region (see \S\ref{sec:CO} and Figure~\ref{fig:CO_ratios}).
\item Regions with the highest CO column densities, tracing where most of the gas mass is located, surround and hug the radio jet feature. This morphology likely traces out the extent of the jet-blown cavity in this region (see \S\ref{sec:column} and Figure~\ref{fig:colco}).
\item CO lines display asymmetric, double peaked line profiles, in regions located at the base of, and within the radio jet feature. This indicates the presence of multiple gas components, at different velocities, consistent with a collision between the jet and molecular gas in these regions (see \S\ref{sec:CO}, Figure~\ref{fig:CO_mom0_spec}, and Table~\ref{table:Co_line_vals}).
\item {The lack of H$_2$CO emission indicates the absence of hot molecular gas in the suspected cavity region (where the molecular gas may have been excavated by the jet in this region; see \S\ref{sec:h2co}, Figures~\ref{fig:H2CO_mom0} \& \ref{fig:radex}, and Table~\ref{table:h2co_line_vals}).}
\item Shock-tracing emission (SiO) is detected in a compact region near the jet impact zone, potentially indicating a weak shock in the molecular gas at this site (see \S\ref{sec:st}, Figure~\ref{fig:st_mom0}, and Table~\ref{table:st_line_vals}).
\end{enumerate}

\begin{figure}
\begin{center}
\includegraphics[width=1\columnwidth]{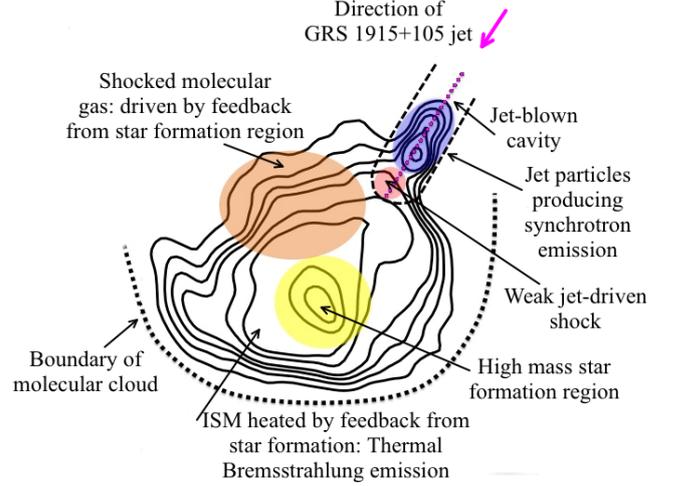}
\caption{\small \label{fig:schem_up} Schematic of the IRAS 19132+1035 region, mapping out the key morphological and emission features detected in our ALMA data. {All regions of interest are colour coded and labelled, and the magenta dotted line indicates the center of the jet-blown cavity.} The contours represent the VLA radio continuum contours; levels of 0.2, 0.3, 0.4, 0.5, 0.75, 1, 2, and 3 ${\rm mJy\,bm}^{-1}$ (see Figure~\ref{fig:grsconts}). Our high resolution ALMA data has allowed us to distinguish between two different feedback mechanisms powering the IRAS 19132+1035 region; the BHXB jet and a high mass star formation region. }
\end{center}
\end{figure}

While all of this {new evidence supports a jet-ISM interaction in the IRAS 19132+1035 region}, we find that this interaction occurs on much smaller scales than postulated by previous works, and does not dominantly power or shape the whole IRAS 19132+1035 region. Rather, our data suggests that in the IRAS 19132+1035 region, we are observing a weaker long-range jet-ISM interaction with a molecular cloud that also hosts a high-mass star forming region. Figure~\ref{fig:schem_up} displays an update on the \cite{kai04} schematic model (shown in Figure~\ref{fig:grs} \textit{right}) of the IRAS 19132+1035 region, based on our new data, where we label key morphological and emission features.

\subsection{Star formation feedback in the IRAS 19132+1035 region}
\label{sec:hmsf}

To estimate the influence that star formation may have had on shaping and powering the IRAS 19132+1035 region, we consider the thermal component of the radio continuum to estimate the properties of a young cluster that would be required to explain the thermal continuum. We calculate the emission measure (EM) of the region to be $1.3\times 10^5~\mathrm{pc~cm^{-6}}$ \citep{toolsra}, based on the peak brightness temperature of the radio emission at $\nu = 5$~GHz ($T_\mathrm{B} = 8.3~\mathrm{K}$; from Table \ref{table:mwemiss}). The projected area of the radio continuum emission region is $\Omega = 1200~\mathrm{arcsec}^2$. If we assume a spherical geometry with $R= (\Omega D^2/\pi)^{1/2}$ at a distance of $D=8.6~\mathrm{kpc}$, the radius of the region is $R=0.81$~pc, and the implied density for a uniform gas is $n_e = 290~\mathrm{cm}^{-3}$.  Assuming a steady state, a pure-hydrogen nebula, and Case B recombination
(using $\alpha_B = 4\times 10^{-13} {\, \rm cm}^3 {\, \rm s}^{-1}; \, T_e=1.2\times 10^4 {\, \rm K}$ \citealt{kai04}),
we require an ionizing photon budget of $Q \sim (\frac{4\pi}{3} R^3 \alpha_B n_e^2) \sim 2\times 10^{48} {\rm s}^{-1}$, which corresponds roughly to the output of a single O8.5V star with mass $M\sim 19~M_{\odot}$ \citep{martins05}. 

Given the coarseness of the estimate, this conclusion is consistent with the analysis of \citet{kraemer03}, who analysed the mid-infrared and 3.6 cm radio continuum fluxes, and determined that the driving source for this object is a B0V star (G45.19 in their study).  Assuming a \citet{kroupa} IMF, this implies the region hosts a small cluster with a total mass of $M_{\star} \sim 200~M_{\odot}$, with some uncertainty owing to the stochastic sampling of the IMF.  The mass of the molecular gas, as seen in the entire ALMA imaging field is $\sim 500~M_{\odot}$, which implies a typical star formation efficiency of $\epsilon_{\mathrm{SF}} = M_\star / (M_{\mathrm{gas}} + M_{\star}) = 0.3$ \citep[cf.,][]{lee16}.

\begin{figure}
\begin{center}
\includegraphics[width=1\columnwidth]{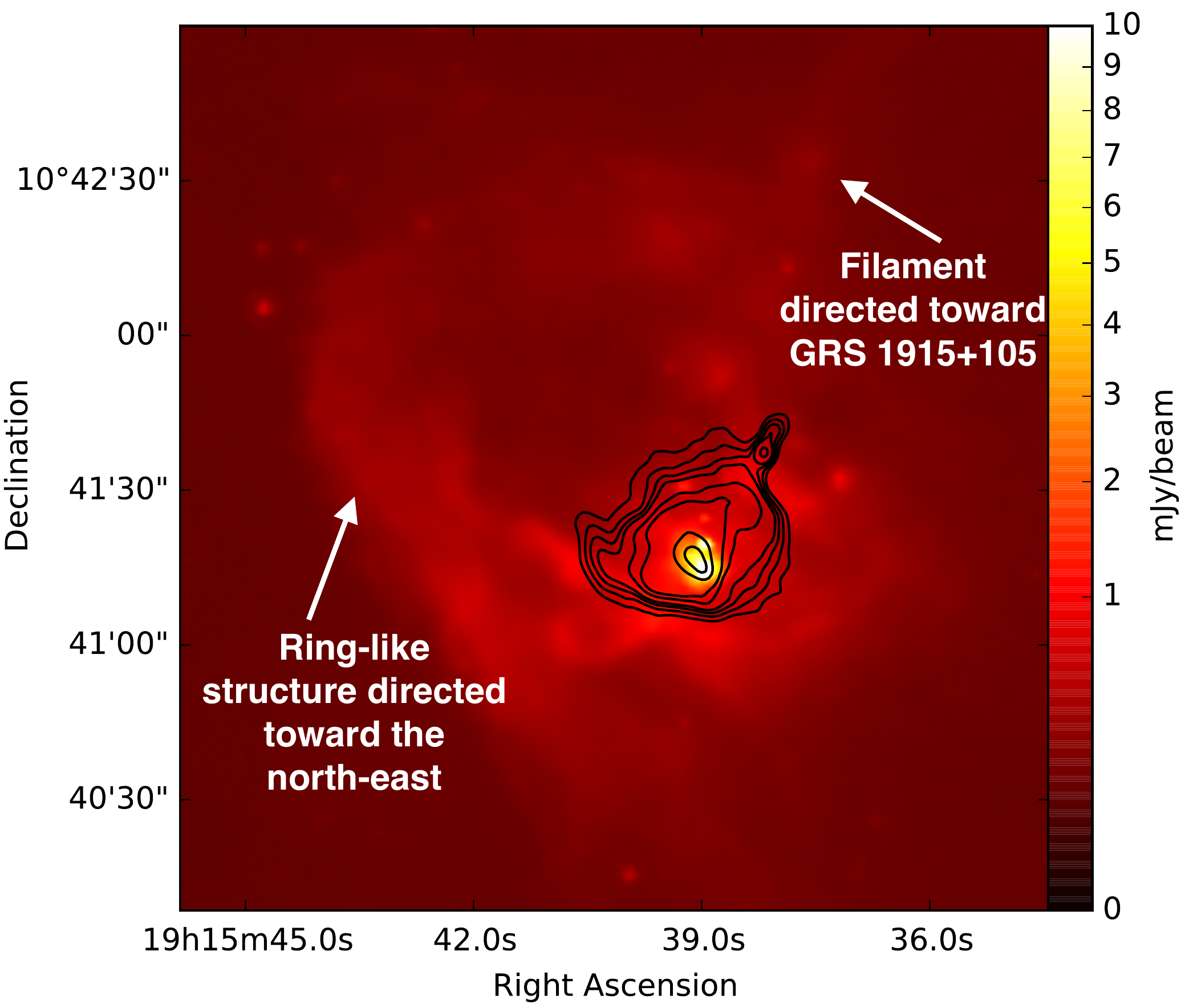}
\caption{\small \label{fig:pacs}. Zoomed out version of the $8\mu m$ Spitzer GLIMPSE continuum image displayed in the bottom left panel of Figure~\ref{fig:grsconts_mw}, where we have labeled the fainter, diffuse emission features observed around IRAS 19132+1035. The contours represent the VLA radio continuum contours; levels of 0.2, 0.3, 0.4, 0.5, 0.75, 1, 2, and 3 ${\rm mJy\,bm}^{-1}$ (see Figure~\ref{fig:grsconts}). The colour bar displays the $8\mu m$ flux density in units of ${\rm mJy\,bm}^{-1}$. The morphology of the diffuse dust emission in this infrared map is consistent with a young cluster in the region driving a blister {\sc Hii} region expanding at the sound speed.}
\end{center}
\end{figure}

Using the same radius for the infrared emission and the dust properties derived in Table \ref{table:greybody}, we calculate the infrared luminosity of the associated cluster to be $L\sim 2.8\times 10^4~L_{\odot}$.  If the cluster luminosity is dominated by the O star \citep[$L\sim7\times 10^4~L_{\odot}$][]{martins05}, this implies $<40\%$ of the luminous output is re-radiated in the infrared.  This finding is consistent with the ``bubble'' morphology seen in the infrared imaging (Figure \ref{fig:pacs}). The size of the larger blowout is $\sim 2$~pc, all consistent with a young cluster driving a blister {\sc Hii} region \citep{gendelev12} expanding at the sound speed in $T\sim 10^4~\mbox{ K}$ gas giving an age of $t \sim 0.2$~Myr.  If we take $E_{\mathrm{SF}} = L_{\mathrm{IR}} t$ as the amount of energy deposited as feedback by star formation in the region, we find $E_{\mathrm{SF}}=7\times 10^{50}~\mathrm{erg}$.  The remainder of the energy is assumed to leave the region as optical/UV light.

In summary, IRAS 19132+1035 is small molecular cloud ($M\sim 500~M_{\odot}$) that hosts a young, medium mass stellar cluster with $M\sim 200~M_{\odot}$ and age $t\sim 0.2~\mathrm{Myr}$.  A single O9 star is sufficient to explain (1) the total infrared luminosity in this region, (2) the thermal radio continuum, and (3) the blowout morphology seen in the mid- and far-infrared imaging.  Such a young region would also be sufficient to explain the weak SiO emission and the gas heating seen in the H$_2$CO emission. Given the apparent youth of the region, the non-thermal emission in the radio continuum is difficult to explain through supernovae and thus represents good evidence for the presence of the jet interaction in the field of view.

\subsection{Jet induced star formation}
Given that we have shown compelling evidence that the GRS 1915+105 jet is colliding with a molecular cloud in IRAS 10132+1035, which houses a high mass star forming region, we must consider the possibility that the jet has triggered the star formation process in this region. Previous studies of the IRAS 19132+1035 region have suggested that, given the compact nature of the molecular gas in the region, the jet may not only have triggered star formation in the region (via compression of the molecular gas), but may also have induced the formation of the original molecular cloud by gradually collecting the interstellar gas in its path \citep{chat01,mir15}. 
In order for either of these processes to occur, the GRS 1915+105 jet must have been active for much longer than we have been able to observe it (i.e.\ prior to 1992). For example, the approximate timescale for a molecular cloud to collapse and begin to form stars is on the order of a few Myr \citep{caz02}.


While the age of GRS 1915+105 (and in turn the timescale of past jet activity) is not well constrained, there are many observable clues that suggest GRS 1915+105 may in fact be a very old stellar system.
Given measured proper motions of the jet ejecta launched from GRS 1915+105 ($\mu_{\rm app}=17.6\,{\rm mas\,day}^{-1}$, $\mu_{\rm rec}=9.0\,{\rm mas\,day}^{-1}$; \citealt{mirbel4}), and the known inclination angle ($i=60\pm5^{\circ}$; \citealt{reid2014}), we estimate the true velocity of the jet ejecta to be
\begin{equation}
\beta\,{\rm cos}(i)=\frac{\mu_{\rm app}-\mu_{\rm rec}}{\mu_{\rm app}+\mu_{\rm rec}}=0.94c.
\end{equation}
When combined with the $42.5\,{\rm pc}$ separation between GRS 1915+105 and IRAS 19132+1035 (i.e.\ 17\arcmin at 8.6 kpc), this implies a travel time of the jet between the two of $>150$ years (allowing for the potential of a decelerating jet), which supports the hypothesis that the source must have been active prior to 1992.
Additionally, the donor star in GRS 1915+105 is known to be a red giant undergoing Roche lobe overflow, meaning the source has likely been actively accreting (and producing a jet) on a timescale at least as long as the time for the donor star to cross the Hertzsprung gap and remain in the giant branch ($\sim 10^{2}-10^{3}$ years; \citealt{mir15}).
Further, GRS 1915+105 is one of a handful of systems with measured 3-dimensional velocities \citep{mj14}. The low measured peculiar velocity of GRS 1915+105 ($22\pm24$  ${\rm km\, s}^{-1}$)\footnote{As this measurement is consistent with zero velocity (within the errors), it cannot be used to derive an age estimate.} indicates that this system likely did not receive a large natal kick at birth (i.e.\ no supernova explosion), but rather may have obtained its current velocity by galactic diffusion (random gravitational perturbations due to collisions with the spiral arms and giant molecular clouds; \citealt{reid2014}). In this case, GRS 1915+105 could have orbited the Galaxy several times, and be as old as a few Gyr \citep{dhaw07,mir16}. 

Even though the predicted age of GRS 1915+105 may be consistent with the timescales for triggered star formation to occur, the thermal dust morphology provides hints that this may not be the case for IRAS 19132+1035. In particular, a ring-like structure directed towards the north-east can be seen in the $8\mu m$ Spitzer GLIMPSE continuum image of the region surrounding IRAS 19132+1035 (Figure~\ref{fig:pacs}). The presence of this structure suggests that the star formation in the IRAS 19132+1035 zone could have been triggered by another bout of star formation to the north-east, rather than the jet from GRS 1915+105, which is observed to come from the north-west direction. However, we note that \cite{mir15} identify a faint filamentary structure in a $160\mu m$ image of the same field (also visible in our $8\mu m$ image), pointing in the direction of GRS 1915+105; these authors suggest this structure could indicate the BHXB jet played at least a minor role in shaping the molecular cloud and governing the star formation activity in this region.

\subsection{Constraints on jet properties}
\label{sec:jetprop}
A common technique in AGN jet studies involves using the interaction sites between the jets and the surrounding medium (i.e.\ the large-scale jet-blown radio lobes) as accurate calorimeters, to estimate the jets' \textit{power} $\times$ \textit{lifetime} products (e.g.\ \citealt{bur59,cas75,kai97,mac07}). This technique can also be applied to jet-ISM interaction sites near Galactic BHXBs in the cases where a jet-blown bubble or cavity is detected (e.g.\ Cyg X-1; \citealt{gallo05,rus07,sell15}).

We follow the self-similar fluid model of \cite{kai97}, which balances the ram pressure of the shocked ISM with the interior pressure exerted by the jet-blown cavity {(see Figure~\ref{fig:schem_up} for a schematic showing the cavity in IRAS 19132+1035, where non-thermal synchrotron emission in the cavity is coloured blue)}. 
Assuming the jet direction remains constant (i.e.\ the jet is not precessing), the jet is colliding with a medium of density $\rho_0$, 
the power being transported by the jets ($Q_{\rm jet}$; averaged over the lifetime of the jets), solely dependent on the properties of the ISM at the interaction site, is represented as,
\begin{equation}
\label{eq:pow}
Q_{\rm jet}=\left(\frac{5}{3}\right)^3\frac{\rho_0}{C_1^5}\,L_j^2\,v^3
\end{equation}
where $L_j$ represents the length of the jet as a function of time ($t$), $C_1$ is a constant dependent on the adiabatic indices of the material in the jet, cavity, and external medium ($\Gamma_j$, $\Gamma_x$, $\Gamma_c$) and the jet opening angle ($\theta$), and $v$ represents the velocity of the shocked gas at the interaction site {(colored red in Figure~\ref{fig:schem_up})}. The full derivation of this jet power expression is provided in Appendix~\ref{sec:cal_app}.

Given the angular distance of IRAS 19132+1035 from GRS 1915+105 {(17\arcmin; see Figure~\ref{fig:grs} \textit{left})}, the length of the jet, $L_j$, can be expressed as \citep{kai04},
\begin{equation}
L_j=1.52\times10^{19}\,\frac{D}{\sin i}\quad{\rm cm},
\end{equation}
where $D$ represents the distance to the source in kpc, and $i$ represents the inclination angle of the jet to our line of sight.

As the source geometry is unknown, to estimate the average density of the medium the jet is colliding with in IRAS 19132+1035, we follow \cite{kai04}, and model the region as a sphere of diameter 36.5\arcsec {(centered at coordinates 19:15:39.087, +10:41:22.652, to cover the extent of the radio continuum feature; see Figures~\ref{fig:grs} and \ref{fig:schem_up})}, at the distance to the source. The volume (in units of ${\rm cm}^{-3}$) of this spherical region is represented as
\begin{equation}
V_{\rm sphere}=8.53\times10^{52}D^3f,
\end{equation}
where $f$ represents a volume filling factor ($f<<1$ would indicate a hollow shell). Using our H$_2$ column density map (see \S\ref{sec:column} and Appendix~\ref{sec:column_app} for details), we estimate the gas mass\footnote{The mass reported here is estimated from the molecular gas contained by the radio continuum feature, while the mass reported in \S\ref{sec:hmsf} is estimated from the molecular gas contained in the whole ALMA field of view. We use the former for these calculations, as this estimate is more representative of the gas mass interacting with, and being displaced by, the BHXB jet.} in this spherical region to be $226~M_{\rm \odot}$.
These measurements yield an estimate of the average density in the region of
\begin{equation}
\rho_0=2.65\times10^{-18}D^{-3}f^{-1}\quad{\rm g\,cm}^{-3}.
\label{eq:rho0}
\end{equation}

Through combining these expressions, we can place constraints on the average power carried by the BHXB jet over its lifetime. We performed Monte Carlo simulations, sampling from the known distributions of distance ($D=8.6^{+2.0}_{-1.6}$ kpc), inclination ($i=60\pm5^\circ$), opening angle (we use a uniform distribution between 0 and $4^\circ$, as \citealt{millerj06} report opening angle constraints of $\theta<4^\circ$), and shocked gas velocity ($v=1.01^{+0.24}_{-0.18}\,{\rm km\,s}^{-1}$; equivalent to the FWHM of the northernmost SiO detection, {coloured red in Figure~\ref{fig:schem_up}}), and setting $f=0.1$ (a reasonable assumption given that the jet is displacing molecular gas), and $\Gamma_j=\Gamma_x=\Gamma_c=5/3$. This procedure yields
an estimate of the energy carried in the GRS 1915+105 jet of $(6.7^{+6.4}_{-6.6})\times10^{47}\,{\rm erg}$, over a lifetime of $29.5_{-8.7}^{+8.6}$ Myr, resulting in a total time-averaged jet power of $(8.4^{+7.7}_{-8.1})\times10^{32}\,{\rm erg\,s}^{-1}$.

Our GRS 1915+105 jet power estimate lies below the distribution of estimated jet powers in the BHXB population ($10^{36}-10^{39}\, {\rm erg\,s}^{-1}$; \citealt{curr14}), and is orders of magnitude lower than estimated from the intensity/duration of the radio flares that accompany transient jet ejections launched by GRS 1915+105 (e.g.,$10^{39}\, {\rm erg\,s}^{-1}$; \citealt{fen99}). 
This discrepancy in estimated jet power could be a result of simplifying assumptions used in both methods leading to over/under estimations of the jet power in either case.
For example, in our calorimetric method we have assumed a constant jet power over the source's lifetime. However, as GRS 1915+105 was only discovered when it entered outburst in 1992, the source must have spent at least some of its pre-1992 lifetime in a significantly less active quiescent state (presumably launching less powerful jets; \citealt{plotkin2015}). Further, in addition to transient ejection events \citep{mirbel4}, GRS 1915+105 has also been observed to undergo extended periods in which a compact jet is present instead \citep{dhaw00}. As these different types of jets are thought to produce different radiative energies, travel at different bulk speeds (resulting in varying kinetic energies), and be active over different timescales \citep{fenbelgal04,fend06,rus13aa}, it is reasonable to expect that the power output from the jets does not remain constant in this source.

We have also assumed a constant external density. Given the $\sim 40$ pc distance between GRS 1915+105 and IRAS 19132+1035, this is unlikely to be the case. While there is no strong evidence that the transient ejections launched from GRS 1915+105 are decelerating on large (arcsec) scales (presumably due to a collision with a denser medium),  deceleration on smaller scales ($<70$ mas; \citealt{millerj07}) cannot be ruled out. This in turn could indicate the presence of a denser medium closer to the source, violating this second assumption.
The unknown geometry of IRAS 19132+1035 is also a source of uncertainty in our calculations.
Equally, estimates of jet power using the radio flaring counterparts to transient ejections are known to fold in many assumptions about unknown jet properties (i.e.\ composition, efficiency, equipartition between jet particles and the magnetic field), which could contribute to over/under estimations of the jet power in this case. Further, given that the transient jet events are rare, and relatively short lived when compared to the compact jets in this source, their high jet power is not likely representative of the average jet power over the source's lifetime.
 
On the other hand, given the {uncertainty} regarding outburst duration and jet duty cycle in GRS 1915+105, our lower time-averaged jet power measurement may instead reflect a jet that carries a much higher energy than we have estimated, but has only been turned on for a relatively small fraction of its total lifetime (rather than a jet with a constant, lower power output). For instance, the discrepancy between our jet power and other estimates suggests that the jet has only been on (i.e.\ GRS 1915$+$105 has roughly been in outburst) for a fraction of $10^{-3}$--$10^{-6}$ of its total lifetime.

Further, it has been postulated that the GRS 1915+105 jet may be precessing \citep{rodmir99,rush10}, which could result in the energy carried by the jet being smeared around a conical path, with only a fraction of the arc traced out by the precessing jet depositing energy into the molecular cloud.  Assuming the precessing jet traces out a cone, with opening angle of $\sim10$ degrees \citep{rodmir99} and edge length of the distance between the sources (42.5 pc), then the power in the jet would be larger by a factor of $\frac{2\pi R_{\rm cone}}{{\rm arc length}}\sim 150$ (for $R_{\rm cone}=7$ pc and ${\rm arclength}=0.3$ pc), resulting in $Q_{\rm jet,precess}\sim 1\times10^{35}\,{\rm erg\,s}^{-1}$.

Moreover, until this point, we have only considered the case where the GRS 1915+105 jet is always pointed in the direction of IRAS 19132+1035. In fact, it is unlikely that the BHXB source and the IRAS 19132+1035 source would be lined up for the $\sim30$ Myr jet lifetime we have estimated above. If the relative velocities between the molecular cloud in IRAS 19132+1035 and the BHXB are different, the jet could be slicing through the molecular cloud for a limited period of time, only depositing energy into the molecular cloud during traversal. We briefly consider this case by exploring a simple model, where we assume an elastic collision between the jet and molecular cloud. In this case, the power in the jet could be estimated by the kinetic energy of the displaced mass, divided by the interaction time, according to
\begin{equation}
Q_{\rm jet,slice}=\frac{\frac{1}{2}\delta M \delta v^2}{\left(\frac{R_{\rm cloud}}{\Delta V}\right)},
\end{equation}
where $\delta M$ represents displaced mass, $\delta v$ represents the velocity of the displaced gas, and $\Delta V$ represents the magnitude of the difference between the velocity vectors of the BHXB and molecular cloud. 
Substituting in values calculated above, $\delta M=226\, M_\odot$, $\delta v=1\,{\rm km\,s}^{-1}$, $R_{\rm cloud}=2$ pc, and $\Delta V=22\,{\rm km\,s}^{-1}$ (equivalent to the BHXB's peculiar velocity\footnote{We note that given the large uncertainties on the peculiar velocity measurement, it is also consistent with zero (22+/-24 km/s).}, where we assume the molecular cloud undergoes pure Galactic rotation), yields $Q_{\rm jet,slice}\sim 4\times10^{32}\,{\rm erg\,s}^{-1}$. This jet power estimate is on par with that estimated from the calorimetric method above.

Overall, from the above calculations, it is clear that the jet is depositing relatively very little energy into the molecular cloud ($\sim10^{47}\,{\rm erg}$), especially when compared to the feedback from the star formation process ($\sim10^{50}\,{\rm erg}$), and thus does not dominantly power the IRAS 19132+1035 region.

Lastly, given the extent of the jet-blown cavity detected in the molecular gas (8 arcsec at 8.6 kpc, equivalent to 0.3 pc) in IRAS 19132+1035, we can also place an estimate on the GRS 1915+105 jet opening angle where,
\begin{equation}
\phi=\arctan(0.3\, {\rm pc}/42.5\, {\rm pc})=0.4^\circ.
\end{equation}
This opening angle estimate is consitent with the upper limit of $<4^\circ$, reported in \citet{millerj06}.

\subsection{Comparison to other known jet-ISM interaction zones}
\label{sec:compare}

Given our uncertainty in how BHXB jet-ISM interactions are likely to manifest in the surrounding ISM, it is of interest to compare the features of other interaction sites with our observations of IRAS 19132+1035. 
Prior to this work, two confirmed jet-ISM interaction sites near BHXBs in our Galaxy, which show both a jet-blown bubble/cavity and evidence for shock excited gas, had been identified; SS 433 and Cyg X-1. Additionally, there are four other BHXBs, in which a jet-ISM interaction has been invoked to explain atypical radio and X-ray emission properties, such as downstream radio/X-ray hot-spots and decelerating jet components; XTE J1550-564, H1743-322, XTE J1752-223, XTE J1908+094.

SS 433 is known to launch precessing jets \citep{mar84,hjon81,fabrika2004}, and is located inside the large-scale W50 radio nebula ($192\times96$ pc at 5.5 kpc distance). This nebula displays a unique ``sea-shell" morphology, consisting of a central spherical component, and two oppositely directed, elongated ``ears", believed to have been inflated by the jets \citep{dubn98}. The eastern ear shows a helical pattern within the radio continuum morphology, thought to mirror the large scale precession of the jets. The western ear (which is believed to be located in a higher density medium) is smaller in scale, brighter at radio frequencies, and displays multiple hot-spots \citep{fuch02} that coincide with X-ray \citep{sah97}, infrared \citep{band87,on00}, and sub-mm molecular emission (CO, HCO$^{+}$; \citealt{huang83,dur00,chat01}). These hot-spots are thought to mark sites where the precessing jet collides with the ambient medium, and the dust and gas are being heated by a shock driven jet-ISM interaction 
\citep{fuch02}. \cite{dubn98} estimated the jets of SS 433 inject an energy of $\sim10^{51}$ erg, over a lifetime of $\sim10^{4}$ years, into the surrounding nebula, resulting in an estimated total jet power of $\sim10^{39}\, {\rm erg\, s}^{-1}$.

Cyg X-1 displays persistent, radio emitting jets \citep{stir01} that are thought to be propagating through the tail end of a dense H{\sc ii} region \citep{rus07,sell15}.  Deep radio-frequency observations of the field surrounding Cyg X-1 revealed a shell-like structure $\sim5$ pc in diameter, aligned with the jet axis of Cyg X-1 \citep{gallo05}. This shell structure was also observed in optical line emission (H$\alpha$, H$\beta$, [O{\sc iii}], [N{\sc ii}], [S{\sc ii}]; \citealt{rus07,sell15}), although no diffuse X-ray emission was detected in the region \citep{sell15}. To date, no molecular emission has been detected in or around the shell structure.
\cite{gallo05} interpreted this shell as a jet-blown bubble, containing thermal plasma, which formed as the result of a strong radiative shock created when the jet impacted the ISM. However, alternative explanations for the shell have been put forward, including a supernova remnant, and a shock wave driven by the strong stellar wind from the high mass companion star in Cyg X-1 \citep{sell15}. \cite{gallo05} inferred that the jets of Cyg X-1 would need to carry a time-averaged power of $\sim9\times10^{35}-10^{37}\,{\rm erg\, s}^{-1}$, over a lifetime of $\sim0.02-0.32$ Myr, to create and maintain the jet-blown bubble structure in the surrounding medium.

XTE J1550-564, H1743-322, XTE J1752-223, and XTE J1908+094 are transient BHXBs, which occasionally enter into bright outburst periods, typically lasting on the order of months.
During these outbursts, all of these systems have been observed to launch transient jet ejecta.
Following these ejection events, compact radio and X-ray hot-spots, aligned with the jet axis, and located at distances ranging from hundreds of mas up to arcseconds from the central source (sub-parsec physical distance scales), have been observed in the fields surrounding these systems \citep{co05,cor02,karr3,yang10,millerjones11,yang11,ratti12,rush17}. Lateral expansion was detected in the radio hot spots of XTE J1752-223 and XTE J1908+094 \citep{yang10,rush17}.
These features have been interpreted as working surfaces, where the jet ejecta collide with the surrounding ISM, creating a moving shock front. Such features are similar to those observed in the neutron star system, Sco X-1 \citep{fom01,foma01}.
Further, deceleration detected in the bulk motion of jet ejecta components launched from XTE J1550-564, H1743-322, and XTE J1752-223, on timescales of tens to hundreds of days, provides additional evidence in favor of the jet colliding with a dense medium surrounding these BHXBs \citep{co05,cor02,yang10,millerjones11}.
In XTE J1550-564 and H1743-322, the two hot spot components (located on both sides of the BHXB) were initially detected at the same angular distance from the central source. This may indicate that the jets are propagating through an evacuated cavity (carved out during a previous outburst by jets or accretion disc winds), and become observable only when they collide with a denser phase of the ISM at the edge of the cavity \citep{co05,hao09}. On the other hand, in XTE J1752-223 and XTE J1908+094, asymmetries between the two detected hot spots (i.e.\ differing brightness, expansion rate, and proper motions) have been attributed to varying ISM densities.

In the majority of the interaction cases, there is evidence that the BHXB jet carves out some form of cavity/bubble in the intervening medium as it propagates away from the central BHXB. An interaction only becomes observable when the jet collides with a much denser portion of the surrounding medium (e.g.\ H{\sc ii} region,  radio nebula, or molecular cloud). However, the morphological and emission features at the jet impact sites, as well as the distances traversed by the jet (sub-parsec to tens of parsecs) before an interaction occurs, vary greatly between the different cases. The wide jet-blown bubble of Cyg X-1 terminates in a high velocity  ($\sim 200\,{\rm km\,s}^{-1}$), radiative bow shock located on parsec scales from the central source, SS 433 displays compact hot spots (located tens of parsecs from the central source) where the precessing jet collides with the nebular structure,  the cavity we detected in IRAS 19132+1035 is narrow and terminates in a weak shock ($\sim 1\,{\rm km\,s}^{-1}$)\footnote{We note that a low shock velocity in this case may not be unexpected. \citealt{brom11} derive an analytical expression for the propagation speed of a jet-driven bow-shock (Table 1 and equation B3). Substituting in our values for density, jet power, opening angle, and source age from \S\ref{sec:jetprop}, results in a propagation speed of $\sim0.5\,{\rm km\,s}^{-1}$ (similar to our measured value). Therefore, it seems that finding a low shock velocity is plausible in this case.} at a $\sim40$ pc separation from the central source, and the transient systems all display compact hot spots (located $<1$ pc from the central source) which appear only following jet ejection episodes. Additionally, the only sites that emit non-thermal synchrotron emission (characteristic of BHXB jets) are IRAS 19132+1035 and the transient systems, 
while the only cases where high-energy X-ray emission is detected are in the impact zones of SS 433 and the transient systems.

The observational differences between these interaction sites could be a result of the differing jet properties among the sources. For instance, the high-velocity bow shock near Cyg X-1 may be characteristic of the interaction of a steady, compact jet with the ISM, while the hot spot morphology may be more characteristic of 
jet ejections launched in SS 433 and the transient systems. 
The significantly higher power carried by the SS 433 jets (and presumably the short-lived ejections from the transient systems), in comparison to the GRS 1915+105 jets, may explain why only these jet impact sites emit in the high-energy X-ray bands. As well, the large-scale precession of the SS 433 jets results in the energy carried by the jet being deposited across multiple impact sites, rather than constantly bombarding a single site (as in IRAS 19132+1035). 
All of the transient sources, which undergo occasional, short-lived outbursts (XTE J1550-564, H1743-322, XTE J1752-223, and XTE J1908+094), display compact interaction sites much closer to the central BHXB than the persistently (or long-duration) out-bursting Cyg X-1, SS 433 and GRS 1915+105 systems (sub-parsec scales vs.\ tens of parsecs). Therefore, the timescale over which the jet is active and the duty cycle of the jets, may also govern the size scale and structure of interaction zones. For instance, we may expect parsec scale cavities/bubbles to only form near persistently accreting systems.

Although, the differing ISM environments that the jets are propagating through (e.g.\ molecular gas in IRAS 19132+1035 vs atomic gas in Cyg X-1, different density gradients in the radio nebula surrounding SS 433 when compared to the {\sc Hii} region that the Cyg X-1 jet is propagating through) may also play a key role in the way that a jet-ISM interaction manifests itself in the surrounding medium.  
Further, contributions from other feedback mechanisms in these regions may explain the differences between interaction sites. For instance, given the possibility that the stellar wind may also be powering the Cyg X-1 bubble, the jet-driven shock velocity may have been overestimated, possibly explaining the stark contrast between the estimated shock velocity in the Cyg X-1 and IRAS 19132+1035 cases.
{A comparison between the observational characteristics of these interaction sites and numerical simulations of jet-ISM interactions in XRBs (e.g., \citealt{per08,bord09,bosch11}) could help disentangle the complex processes that are driving such interactions.
Performing numerical simulations of the jet-ISM interaction in IRAS 19132+1035 is beyond the scope of this work, but will be explored in future work.}

So far we have only discussed interactions from black hole systems in our own Galaxy. However, several cases of ionized nebula surrounding ultra-luminous X-ray (ULX) sources (i.e.\ ULX bubbles\footnote{Although we note that a number of these bubble structures have been detected around sources that show X-ray luminosity lower than typical ULXs (i.e., $L_{\rm X}<1\times10^{39}\,{\rm erg\,s}^{-1}$). In these cases, it is likely that either the central X-rays are being obscured, or there has been a change in X-ray luminosity over time. An example of such a source is NGC 7793-S26 \citep{sor10}.}), have been detected in other nearby galaxies (e.g.\ \citealt{rob03,feng08,sor10}). These large-scale (>100 pc) structures are thought to either be shock-ionized (e.g.\ NGC 7793 S26; \citealt{pak10}), purely photo-ionized (e.g.\ NGC 5408 X-1; \citealt{grise12}), or
a combination of the two, where shock-ionized and photo-ionized gas co-exists in the bubble structures. In this last case, photo-ionization occurs from ionizing photons generated by the forward shock (e.g., M83 MQ1 \citealt{soria14}).
The shock-ionized cases can be inflated by different types of outflows; a collimated jet (e.g.\ Holmberg II X-1; \citealt{cseh14}, NGC 7793 S26; \citealt{pak10}, IC342 X-1; \citealt{cseh12}) or winds (e.g.\ NGC 1313 X-2; \citealt{grise08}). While the ULX bubbles are much larger in scale than most of the Galactic interaction zones, these ULX interactions display striking similarities to many Galactic cases, such as radio and X-ray hotspots characteristic of jet impact sites (similar to SS 433 and the transient systems; e.g., NGC 7793 S26), optically thin radio synchrotron emission (similar to IRAS 19132+1035; e.g., Holmberg II X-1), and optical line emission characteristic of shock excited gas (similar to Cyg X-1; e.g., M83 MQ1).
Further, similar to the black hole systems in our own Galaxy, it appears as though the way in which an ULX interaction manifests is also highly dependent on the properties of the central source and the local environmental conditions.

\subsection{Jet molecular cloud interaction probabilities}
\label{sec:prob}
In this work we have shown that the jet launched from GRS 1915+105 is colliding with a molecular cloud in IRAS 19132+1035. As such, we wish to estimate the likelihood of detecting more jet molecular cloud collisions in our Galaxy.
This interaction probability can be found by estimating the chance that a given BHXB jet is found within the volume where molecular clouds are formed in our Galaxy. 

We begin by modelling a population of molecular clouds using the principal observation that molecular clouds have constant, average column density regardless of mass or radius; $\Sigma_0 \sim 10^{2}~M_{\odot}~\mathrm{pc}^{-2}$ (e.g.\ \citealt{sol87,hey09}).  The mass distribution of this population will be expressed as a power law distribution with a truncation at the upper mass end $M_u=10^6~M_{\odot}$ (Colombo et al., in preparation).  Assuming that clouds below a mass of $M_l=10^2~M_{\odot}$ are diffuse and will not show significant jet-ISM interactions, the mass distribution can be represented as,
\begin{equation}
\frac{dN}{dM} = \frac{K}{M_u}\left(\frac{M}{M_u}\right)^{\alpha}.
\label{massd}
\end{equation}
Here $\alpha = -1.8$, and $K$ is a normalization constant, such that the integrated mass of molecular clouds yields the total molecular mass in the Galaxy, $M_{\mathrm{H_2}}=10^9~M_{\odot}$ \citep{wol03}. Integrating from $M_l$ to $M_u$ yields $K=200$.  

To transform this mass distribution to a radius distribution, we use the definition $M = \Sigma_0 \pi R^2$. Here $\Sigma_0$ is constant (as defined above) and $R$ represents cloud radius, therefore we can write $dM = 2\pi \Sigma_0 R dR$.
Substituting this result into Equation~\ref{massd} yields,

\begin{equation}
\frac{dN}{dR} = \frac{2K}{R_u}\left(\frac{R}{R_u}\right)^{\beta}.
\end{equation}
where the index $\beta = 2\alpha+1 = -2.6$, $\Sigma_0$ is as above, $R_l = 0.5\mbox{ pc}$, and the radius of the cloud at the upper mass cutoff $R_u = 50~\mathrm{ pc}$.

We define the volume of the Galaxy in which there can be a jet-cloud interaction, as the volume inside a cloud or within a jet length, $R_J$, of a cloud, such that,
\begin{equation}
V_{\mathrm{int}} = \int_{R_l}^{R_u} \frac{4\pi}{3} (R + R_J)^3 \frac{dN}{dR} dR.
\end{equation}
Although, we note that this is an overestimate, since jets at a distances larger than the cloud radius can be oriented such that they will not hit the cloud. 
For compactness in the computation, we set $u \equiv R/R_u$ and $r = R_l/R_u$, yielding,

\begin{align}
V_{\mathrm{int}} &=\frac{8\pi K}{3}\left[
R_u^3 \left(\left.\frac{u^{4+\beta}}{4+\beta}\right|^1_{r}\right) + 
R_u^2 R_J \left(\left.\frac{u^{3+\beta}}{3+\beta}\right|^1_{r}\right)\right.\\\nonumber
&\left.+ R_u^1 R_J^2 \left(\left.\frac{u^{2+\beta}}{2+\beta}\right|^1_{r}\right)+R_J^3 \left(\left.\frac{u^{1+\beta}}{1+\beta}\right|^1_{r}\right)\right] \\ \nonumber
\end{align}

We will take $R_J = 10\mbox{ pc}$. The first term corresponds to the volume within the clouds, for which an interaction is likely. Thus, we obtain a volume of molecular clouds in the Galaxy of $4\times 10^8~\mathrm{ pc}^{3}$. The remaining three terms each contribute $\sim2\times 10^8~\mathrm{ pc}^{3}$, suggesting that, for this jet length, jet-ISM interactions for BHXBs outside clouds, are about as likely as interactions from within the clouds. The total volume in the interaction region is $V_{\mathrm{int}}=10^9\mbox{ pc}^3$.

Secondly, we calculate the volume over which molecular clouds are found in the Galaxy ($V_{\mathrm{mol}}$).  Using the model of \citet{wol03} as a basis, we approximate the molecular medium of the Galaxy as a hollow cylinder with inner radius of $\mathcal{R}_{\mathrm{inner}} = 3$~kpc and $\mathcal{R}_{\mathrm{outer}} = 10$~kpc and a (full) thickness of $h = 100$~pc. For this model, we find $V_{\mathrm{mol}}=3\times 10^{10}\mbox{ pc}^3$.  

Therefore, over the volume where molecular clouds are found, the filling fraction of the interaction volume is equivalent to,
$V_{\mathrm{int}} / V_{\mathrm{mol}} = 3\%$. The interaction probability per X-ray binary where molecular clouds are found is then given by this filling fraction.  Since most low-mass X-ray binaries, which accrete from older stars, are not co-located where molecular clouds are typically found, the above interaction probability likely applies only to low-mass X-ray binaries close to the Galactic plane ($|b| \lesssim0.5\arcdeg$; GRS 1915$+$105 is at $b=-0.2191\arcdeg$). As high mass X-ray binaries by (typical) definition are accreting from an O or B star, they may be preferentially located near molecular clouds and have a higher interaction probability than 3\%.

\subsection{Distance considerations}
\label{sec:dist}

Our detection of an interaction between the GRS 1915+105 jet and the IRAS 19132+1035 region can allow us to place further constraints on the distance to both objects.
The distance to GRS 1915+105 has been estimated from model-independent geometric parallax measurements to be $8.6^{+2.0}_{-1.6}$ kpc, while the kinematic distance to IRAS 19132+1035 is estimated to be $6.0\pm1.4$ kpc. However, for a jet-ISM interaction to be occurring, both GRS 1915+105 and IRAS 19132+1035 must be located at the same distance (within $\sim 40$ pc). While it is difficult to pinpoint the true common distance, given that the current distance constraints are consistent within their one sigma errors, a common distance that lies between the two estimates could satisfy such a condition. 
Note that throughout this paper we have chosen to use the GRS 1915+105 parallax distance in all our calculations as it is model-independent (see \citealt{reid2014b} for a comparison of kinematic and parallax distances). However, if the true distance is indeed closer, this affects our constraints on jet energetics and geometry, as well as the estimated peculiar velocity for GRS 1915+105. For example, assuming a common distance of 7 kpc, we estimate a slightly more energetic jet ($\sim7.3\times10^{47}$ erg), with a shorter lifetime ($\sim$24 Myr), and larger opening angle ($\sim0.5^\circ$), as well as an increased peculiar velocity for GRS 1915+105 ($\sim46\,{\rm km\,s}^{-1}$).

\subsection{IRAS 19124+1106: the second candidate interaction zone near GRS 1915+105}
\label{sec:other_src}

In addition to IRAS 19132+1035, \citealt{rodmir98} also identified another candidate interaction zone to the north of GRS 1915+105; IRAS 19124+1106 (see Figure~\ref{fig:grs}). This second candidate zone is located at a remarkably similar distance from the central BHXB as IRAS 19132+1035, and is also aligned with the position angle of the jet from GRS 1915+105. IRAS 19124+1106 displays a flat radio spectrum, consistent with thermal Bremsstralung emission, and a cometary radio continuum morphology, commonly observed from H{\sc ii} regions. No non-thermal radio emission features are observed in IRAS 19124+1106. Strong molecular emission from the $^{12}$CO (J=2-1), $^{13}$CO (J=2-1), H$^{13}$CO$^{+}$ (J=1-0), and CS (J=2-1) transitions were observed in this region, while no SiO transitions were detected \citep{chat01}. Given the similarities between the emission properties of IRAS 19124+1106 and the star forming zone in IRAS 19132+1035, it seems plausible that IRAS 19124+1106 also contains a star forming molecular cloud. However, the lack of non-thermal radio emission features, as well as the low likelihood that both the approaching and receding components of the GRS 1915+105 jet happen to line up with a star forming molecular cloud, suggests a low probability that IRAS 19124+1106 is also the site of an interaction. As we have shown that sensitive, high resolution observations were needed to observe the jet interaction in IRAS 19132+1035, the same treatment is likely required to draw any further conclusions about the true nature of the IRAS 19124+1106 region.

\subsection{Distinguishing between BHXB jet driven feedback and high mass star formation}
In this work, we have shown that there are two main sources of feedback powering the IRAS 19132+1035 region; the BHXB jet and high-mass star formation.
This suggests that the star formation process may act as a contaminant when probing other BHXB jet-ISM interaction sites in our Galaxy. In particular, as star formation is also known to drive fast outflows, which are likely to generate shocks upon impact with the surrounding gas, it may be difficult to distinguish between BHXB jet-driven feedback and high-mass star formation in terms of the signatures they leave behind in the molecular gas. When comparing molecular emission originating from the BHXB jet-driven feedback zone to that originating from the star formation feedback zone in IRAS 19132+1035, we find that while the SiO shock-tracing emission typically showed similar peak intensities, the emission in the BHXB jet driven feedback zone displayed much narrower line widths, over a more compact emission region. While this could suggest that these properties reflect differences between jet-driven and star formation-driven feedback zones, we require a larger sample of observations of multiple systems to fully understand the differences (if any) that exist between the molecular line properties of these different feedback zones.

\section{Summary}
\label{sec:sum}
In this paper, we present the results of our ALMA observations of IRAS 19132+1035. This region was first identified by \cite{rodmir98} as a potential interaction zone between the jet launched from the BHXB GRS 1915+105 and the surrounding ISM, based on its location and unique radio continuum morphology. However, despite several follow-up observing campaigns, no definitive evidence had been found to confirm this hypothesis, with compelling arguments on either side. 

As molecular lines are excellent diagnostics of shock energetics and ISM excitation, we used these ALMA observations to map the molecular line emission in the IRAS 19132+1035 region, aiming to resolve this long-standing question. We detect emission from the $^{12}$CO [J=2-1], $^{13}$CO [$\nu=0$, J=2-1], C$^{18}$O [J=2-1], ${\rm H}_{2}{\rm CO}$ [$J=3_{0,3}-2_{0,2}$], ${\rm H}_{2}{\rm CO}$ [$J=3_{2,2}-2_{2,1}$], ${\rm H}_{2}{\rm CO}$ [$J=3_{2,1}-2_{2,0}$], SiO [$\nu=0$, J=5-4], CH$_3$OH [$J=4_{2,2}-3_{1,2}$], and CS [$\nu=0$, J=5-4] molecular lines.

Given this molecular emission, we identify several new lines of compelling evidence supporting a connection between the IRAS 19132+1035 region and the GRS 1915+105 jet. In particular,  
the morphological, spectral, and kinematic properties of the detected molecular emission indicate the presence of a jet-blown cavity and weak jet-driven shock at the impact site. However, contrary to the scenario put forward by previous work, we find that feedback from the BHXB jet does not dominantly power or shape the IRAS 19132+1035 region. Rather, a high mass star formation region, housed inside a molecular cloud, is heating the dust and gas in the region, and the jet appears to be simply colliding with this molecular cloud. All the thermal radio and infrared emission from the IRAS 19132+1035 region can be explained by a young, medium mass ($\sim200 M_\odot$) stellar cluster with an age of 0.2 Myr, where it is unlikely that the BHXB jet significantly contributed to shaping the molecular cloud and/or triggering the star formation process in the IRAS 19132+1035 region.

Through considering the properties of the detected cavity and displaced molecular gas in the region, we estimate properties of the GRS 1915+105 jets. Following the self-similar fluid model of \cite{kai97}, we used the detected cavity as a calorimeter to calculate the energy carried in the GRS 1915+105 jet of $(6.7^{+6.4}_{-6.6})\times10^{47}\,{\rm erg}$, over a lifetime of $29.5_{-8.7}^{+8.6}$ Myr, resulting in a total time-averaged jet power of $(8.4^{+7.7}_{-8.1})\times10^{32}\,{\rm erg\,s}^{-1}$ (although this may be up to a factor of $150$ higher if the GRS 1915+105 jet is undergoing small-scale precession). These estimates reiterate that the BHXB jet is depositing relatively very little energy into the molecular cloud, when compared to feedback from the star formation process ($\sim10^{50}$ erg).

Upon comparing the characteristics of the IRAS 19132+1035 interaction zone with other jet-ISM interaction sites in our Galaxy and other nearby galaxies, we find that the morphological and emission features at the jet impact sites can vary substantially between the interaction zones driven by jets launched from different systems. {We find that the way in which a BHXB jet-ISM interaction manifests itself appears to be highly dependent on the jet properties (e.g.\ jet power, duty cycle) and local environment conditions (e.g.\ composition, density, other sources of feedback), similar to what has been observed at jet interaction sites surrounding other astrophysical systems (e.g., AGN, ULXs).}

Overall, our analysis demonstrates that molecular lines are excellent diagnostics of the physical conditions in jet-ISM interaction zones near BHXBs. With the molecular tracers toolbox we have developed here (along with our imaging and analysis techniques), we have opened up a new way to conclusively identify more of these highly sought after interaction sites, and use their ISM conditions to probe jet properties across the BHXB population.


\section*{Acknowledgements} 
We thank the anonymous referee for constructive feedback that improved the manuscript.
The authors wish to thank Gerald Schieven for his help in setting up and configuring the ALMA observations presented in this work.
AJT and PF offer a special thanks to Eric Koch for sharing his knowledge and experience with \textsc{casa}, and for many helpful discussions on spectral line imaging, molecular transitions, and star formation. AJT is supported by an Natural Sciences and Engineering Research Council of Canada (NSERC) Post-Graduate Doctoral Scholarship (PGSD2-490318-2016). AJT, GRS, and EWR are supported by NSERC Discovery Grants. JCAMJ is the recipient of an Australian Research Council Future Fellowship (FT140101082). This paper makes use of the following ALMA data: ADS/JAO.ALMA\#2015.1.00976.S. ALMA is a partnership of ESO (representing its member states), NSF (USA) and NINS (Japan), together with NRC (Canada), NSC and ASIAA (Taiwan), and KASI (Republic of Korea), in cooperation with the Republic of Chile. The Joint ALMA Observatory is operated by ESO, AUI/NRAO and NAOJ. The National Radio Astronomy Observatory is a facility of the National Science Foundation operated under cooperative agreement by Associated Universities, Inc. Herschel is an ESA space observatory with science instruments provided by European-led Principal Investigator consortia and with important participation from NASA. This work makes use of data products from the Wide-field Infrared Survey Explorer, which is a joint project of the University of California, Los Angeles, and the Jet Propulsion Laboratory/California Institute of Technology, funded by the National Aeronautics and Space Administration. This work is also based, in part, on observations made with the UKIRT Infrared Deep Sky Survey, as well as the Spitzer Space Telescope, which is operated by the Jet Propulsion Laboratory, California Institute of Technology under a contract with NASA. In this work we make extensive use of the radio astro tools python packages, including \textsc{SpectralCube},  \textsc{pvextractor}, and \textsc{radio\_beam}, as well as the \textsc{pyspeckit} python package.



\bibliography{ABrefList}



\appendix
\section{ALMA Mosaic Field and Noise Map}
\label{sec:appen}
\begin{figure}
\centering
\includegraphics[width=1\columnwidth]{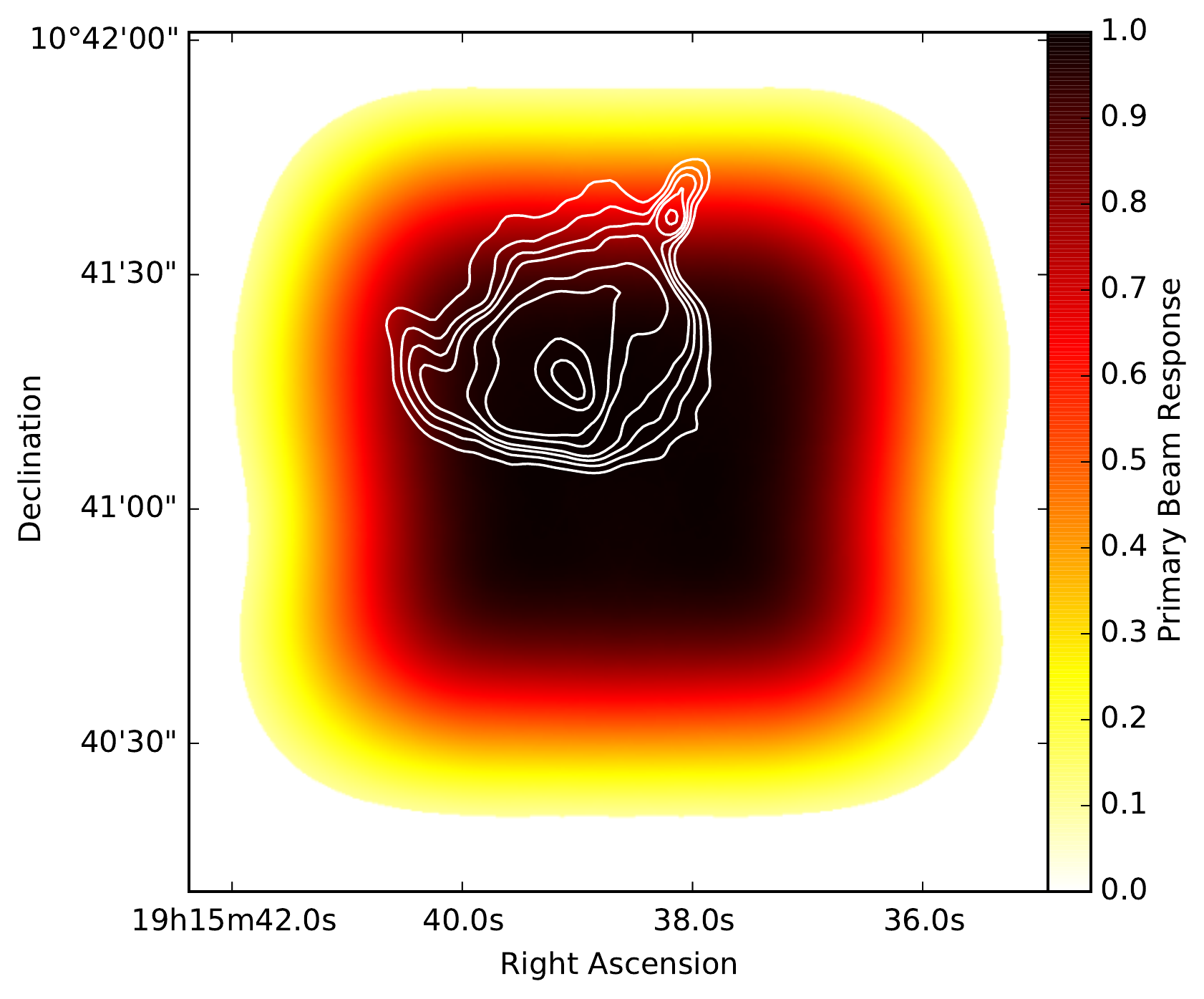}
\caption{\small \label{fig:pbnoise} ALMA primary beam noise map. The contours are the VLA radio frequency contours, as seen in Figure~\ref{fig:grsconts}; 0.2, 0.3, 0.4, 0.5, 0.75, 1, 2, and 3 ${\rm mJy\,bm}^{-1}$. The colour bar indicates the primary beam response. }
\end{figure}

\begin{figure}
\centering
\includegraphics[width=0.9\columnwidth]{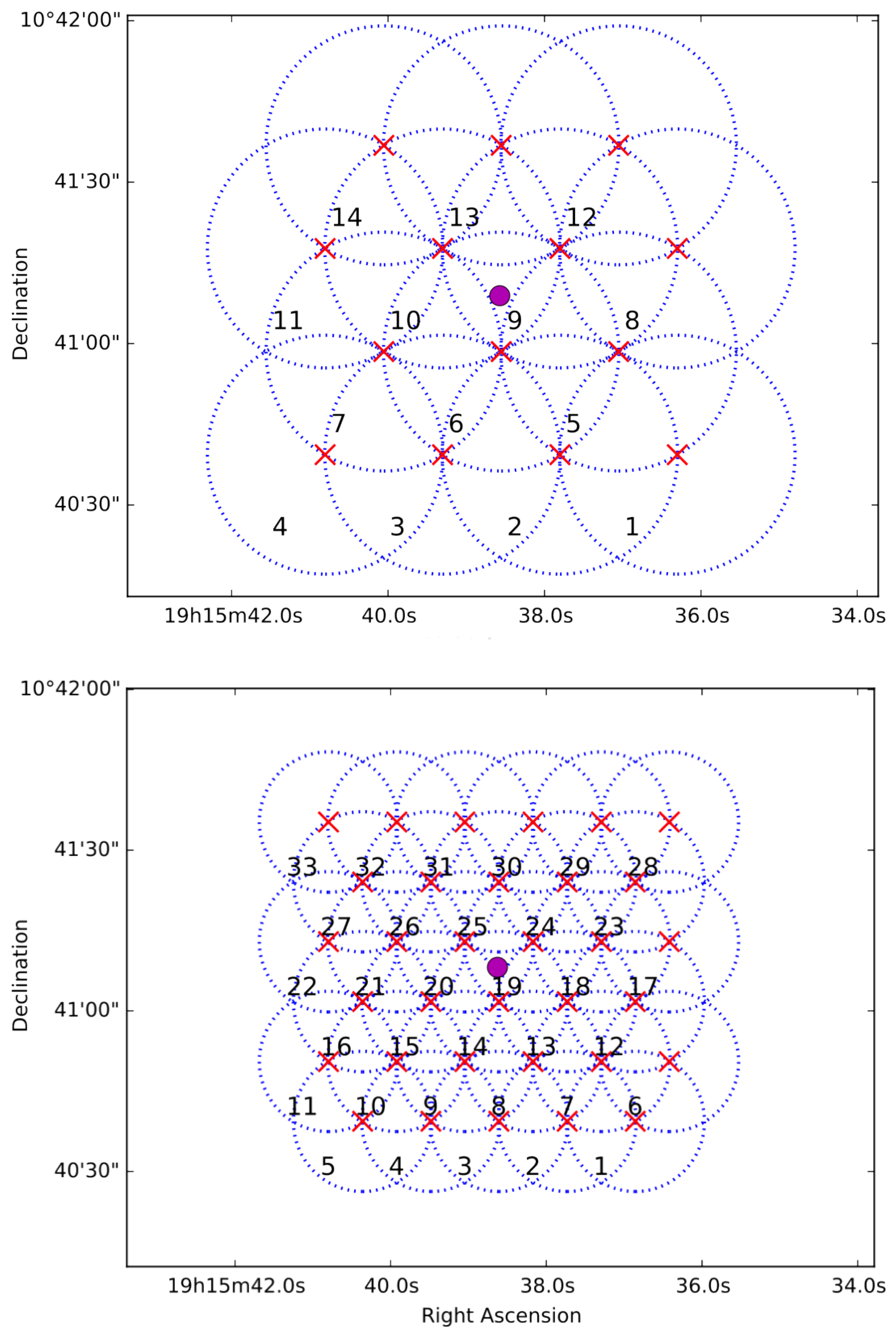}
\caption{\small \label{fig:pbpoint} ALMA ACA 7 m array ({\it top panel}) and 12 m array ({\it bottom panel}) mosaic maps. Each individual pointing (labelled with a number) is represented by a blue circle, with the centre coordinates marked by a red X. The magenta point marks the coordinates of the peak of the radio continuum in IRAS 19132+1035 reported in \citealt{chat01}.}
\end{figure}

We set up our ALMA observations to observe a large rectangular mosaic field, of dimensions 75\arcsec $\times$ 60\arcsec, centred on the coordinates of the peak of the radio continuum, as reported in \citealt{chat01} (J2000 RA 19:15:39.1300, Dec 10:41:17.100). This field was chosen to cover the radio continuum feature, as well as regions to the south of the radio continuum. We did this, as based on previous work, we expected a complex molecular interaction around the suspected bow shock feature, located at the southern most edge of the radio continuum (see Figure~\ref{fig:grs}). A primary beam noise map of the mosaic field is displayed in Figure~\ref{fig:pbnoise}, and pointing maps are displayed in Figure~\ref{fig:pbpoint}.




\section{Column density calculation}
\label{sec:column_app}

We follow the procedure of \cite{toolsra}, assuming that the ${\rm C}^{18}{\rm O}$ emission is optically thin, in LTE, and has an excitation temperature, $T_{\rm ex}$, equivalent to that of the optically thick $^{12}$CO emission \citep{pineda08}. 

The intensity of line emission can be represented as
\begin{equation}
I_{\rm line}=(S-I_0)(1-{\rm exp}[-\tau])
\end{equation}
where $S$ represents the source function, $\tau$ represents the optical depth, and $I_0$ represents the intensity of the initial background radiation field.

Assuming that both the $S$ and $I_0$ can be represented by black-bodies at $T_{\rm ex}$ and $T_{\rm bg}=2.73$ K, respectively, the radiation temperature is represented as
\begin{eqnarray}
\label{eq:tr}
T_R&=&I_{\rm line}\frac{c^2}{2\nu^2k}\\ \nonumber
&=&(S-I_0)(1-e^{-\tau})\frac{c^2}{2\nu^2k}\\ \nonumber
&=& \frac{h\nu}{k}(1-e^{-\tau})\left(\frac{1}{{\rm exp[h\nu/kT_{\rm ex}]}-1}-\frac{1}{{\rm exp[h\nu/kT_{\rm bg}]}-1}\right)\nonumber
\end{eqnarray}

Defining $T_{\rm max,12}$ as the main beam brightness temperature at the peak of the $^{12}$CO line, and assuming that $^{12}$CO is optically thick ($\tau\rightarrow\infty$), Equation~\ref{eq:tr} can be re-arranged to yield
\begin{equation}
T_{\rm ex}=\frac{C_{12}}{\ln\left(1+\frac{C_{12}}{T_{\rm max,12}+C_{12}C_{\rm bg}}\right)}
\end{equation}
where $C_{12}=\frac{h\nu_{12}}{k}$, and $C_{\rm bg}=\frac{1}{{\rm exp}[C_{12}/T_{\rm bg}]-1}$.

Defining $C_{18}=\frac{h\nu_{18}}{k}$, and assuming the excitation temperature of ${\rm C}^{18}{\rm O}$ is equal to that of the $^{12}$CO, the optical depth at a given line-of-sight velocity $V$ can be found by re-arranging Equation~\ref{eq:tr},
\small
\begin{equation}
\tau(V)=-\ln\left(1-\frac{kT(V)}{h\nu_{18}}\left[\frac{1}{{\rm exp}[C_{18}/T_{\rm ex}]-1}-\frac{1}{{\rm exp}[C_{18}/T_{\rm bg}]-1}\right]^{-1}\right).
\label{eq:optdepth}
\end{equation}
\normalsize
Given the optical depth, we can then find the optical column density in the $J$th state from the definitions of the optical depth \citep[see][]{toolsra}:
\begin{eqnarray}
N(J) &=& \frac{8\pi \nu_0^2}{c^2} \frac{g_l}{g_u} \frac{1}{A_{ul}} \left[1-\exp\left(-\frac{h\nu_0}{k T_{\mathrm{ex}}}\right)\right]^{-1} \int \tau d\nu,\\
&=& \frac{8\pi \nu_0^3}{c^3} \frac{g_l}{g_u} \frac{1}{A_{ul}} \left[1-\exp\left(-\frac{h\nu_0}{k T_{\mathrm{ex}}}\right)\right]^{-1} \int \tau dV,
\end{eqnarray}
where the integral is carried out over the line width. We infer the optical depth from Equation \ref{eq:optdepth}.  For C$^{18}$O($2\to 1$), $A_{ul}=1.165\times10^{-11}\mu^2\nu^3\frac{(J+1)}{2J+3}$ for the $J+1\rightarrow J$ transition, and $\mu^2=(0.122D)^2$ for CO.

The total column density is found from summing over all the energy levels of the molecule. Assuming LTE, for a CO molecule ($J+1\rightarrow J$ rotational transition), with a population characterized by a single temperature, $T_{\rm ex}$, the total column density is represented as
\begin{equation}
N_{\rm total}=N(J)\frac{Z}{g_J}{\rm exp}\left[\frac{hB_eJ(J+1)}{kT_{\rm ex}}\right],
\end{equation}
where $g_J=2J+1$ for CO, $k/hB_e=1/2.65\mathrm{~K^{-1}}$, and the partition function is given by,
\begin{equation}
Z=\sum_{J=0}^{\infty}(2J+1){\rm exp}\left[\frac{-hB_eJ(J+1)}{kT_{\rm ex}}\right].
\end{equation}

\section{Calorimetry Method}
\label{sec:cal_app}

Following \cite{kai97}, assuming the jet direction remains constant (i.e.\ the jet is not precessing), the jet is colliding with a medium of density $\rho_0$, 
and that the power is being transported by the jets at a constant rate ($Q_{\rm jet}$; averaged over the lifetime of the jets), the length of the jet as a function of time ($t$) is given by
\begin{equation}
L_j=C_1\left(\frac{t}{\tau}\right)^{\frac{3}{5-\beta}},
\label{eq:Lj}
\end{equation}
where the characteristic timescale $\tau=(\rho_0/Q_{\rm jet})^{1/3}$, and the constants
\begin{equation}
C_1=\left(\frac{C_2}{C_3\theta^2}\frac{(\Gamma_x+1)(\Gamma_c-1)(5-\beta)^3}{18\left[9\{\Gamma_c+(\Gamma_c-1)\frac{C_2}{4\theta^2}\}-4-\beta\right]}\right)^{\frac{1}{(5-\beta)}},
\label{eq:c1}
\end{equation}

\begin{equation}
C_2=\left(\frac{(\Gamma_c-1)(\Gamma_j-1)}{4\Gamma_c}+1\right)^{\frac{\Gamma_c}{(\Gamma_c-1)}}\frac{(\Gamma_j+1)}{(\Gamma_j-1)},
\label{eq:c2}
\end{equation}

\begin{equation}
C_3=\frac{\pi}{4R_{\rm ax}^2}.
\label{eq:c3}
\end{equation}
Here $\theta$ represents the jet opening angle (in radians), $R_{\rm ax}=\sqrt{\frac{1}{4}\frac{C_2}{\theta^2}}$ represents the axial ratio of the jet-blown cavity, and $\Gamma_j$, $\Gamma_x$, and $\Gamma_c$ represent the adiabatic indices of the material in the jet, jet-blown cavity, and external medium with which the jet is colliding, respectively.

The power that the jet would need to carry (averaged over its lifetime, $t$), to create and maintain the jet-blown cavity structure in the surrounding ISM, can be found from re-arranging Equation~\ref{eq:Lj},
\begin{equation}
Q_{\rm jet}=\rho_0{\left(\frac{L_j}{C_1}\right)}^{5} t^{-3}.
\label{eq:Qjet}
\end{equation}
where we have set $\beta=0$ for a constant density medium.

To estimate the jet lifetime, we first take the derivative of Equation~\ref{eq:Lj} (again setting $\beta=0$),
\begin{equation}
\frac{dL_j}{dt}=\frac{3}{5}C_1\left(\frac{Q_{\rm jet}}{\rho_0}\right)^{\frac{1}{5}}t^{-\frac{2}{5}},
\label{eq:dLjdt}
\end{equation}
then combine Equations \ref{eq:Lj} and \ref{eq:dLjdt}, to yield,
\begin{equation}
t=\frac{3}{5}\left(\frac{L_j}{v}\right),
\label{eq:lifetime}
\end{equation}
where $v=dL_j/dt$ represents the velocity of the shocked gas at the interaction site.

Substituting Equation \ref{eq:lifetime} into Equation \ref{eq:Qjet} yields an expression for the power carried by the jet, solely dependent on the properties of the ISM at the interaction site,
\begin{equation}
Q_{\rm jet}=\left(\frac{5}{3}\right)^3\frac{\rho_0}{C_1^5}\,L_j^2\,v^3
\end{equation}



\bsp	
\label{lastpage}
\end{document}